\documentclass[preprint]{revtex4}

\usepackage{color}
\usepackage{graphicx}
\usepackage{amsmath}
\usepackage{amssymb}

\begin{document}

\title{
Mean field theory of weakly-interacting Rydberg polaritons in the EIT system based on the nearest-neighbor distribution}

\author{Shih-Si Hsiao$^1$} 
\author{Ko-Tang Chen$^1$} 
\author{Ite A. Yu$^{1,2,}$}\email{yu@phys.nthu.edu.tw}

\affiliation{
$^{1}$Department of Physics, National Tsing Hua University, Hsinchu 30013, Taiwan \\
$^{2}$Center for Quantum Technology, Hsinchu 30013, Taiwan
}


\begin{abstract}
The combination of high optical nonlinearity in the electromagnetically induced transparency (EIT) effect and strong electric dipole-dipole interaction (DDI) among the Rydberg-state atoms can lead to important applications in quantum information processing and many-body physics. One can utilize the Rydberg-EIT system in the strongly-interacting regime to mediate photon-photon interaction or qubit-qubit operation. One can also employ the Rydberg-EIT system in the weakly-interacting regime to study the Bose-Einstein condensation of Rydberg polaritons. Most of the present theoretical models dealt with the strongly-interacting cases. Here, we consider the weakly-interacting regime and develop a mean field model based on the nearest-neighbor distribution. Using the mean field model, we further derive the analytical formulas for the attenuation coefficient and phase shift of the output probe field. The predictions from the formulas are consistent with the experimental data in the weakly-interacting regime, verifying the validity of our model. As the DDI-induced phase shift and attenuation can be seen as the consequences of elastic and inelastic collisions among particles, this work provides a very useful tool for conceiving ideas relevant to the EIT system of weakly-interacting Rydberg polaritons, and for evaluating experimental feasibility.
\end{abstract}

\maketitle
\newcommand{\FigOne}{
	\begin{figure}[t]
	\center{\includegraphics[width=0.65\columnwidth]{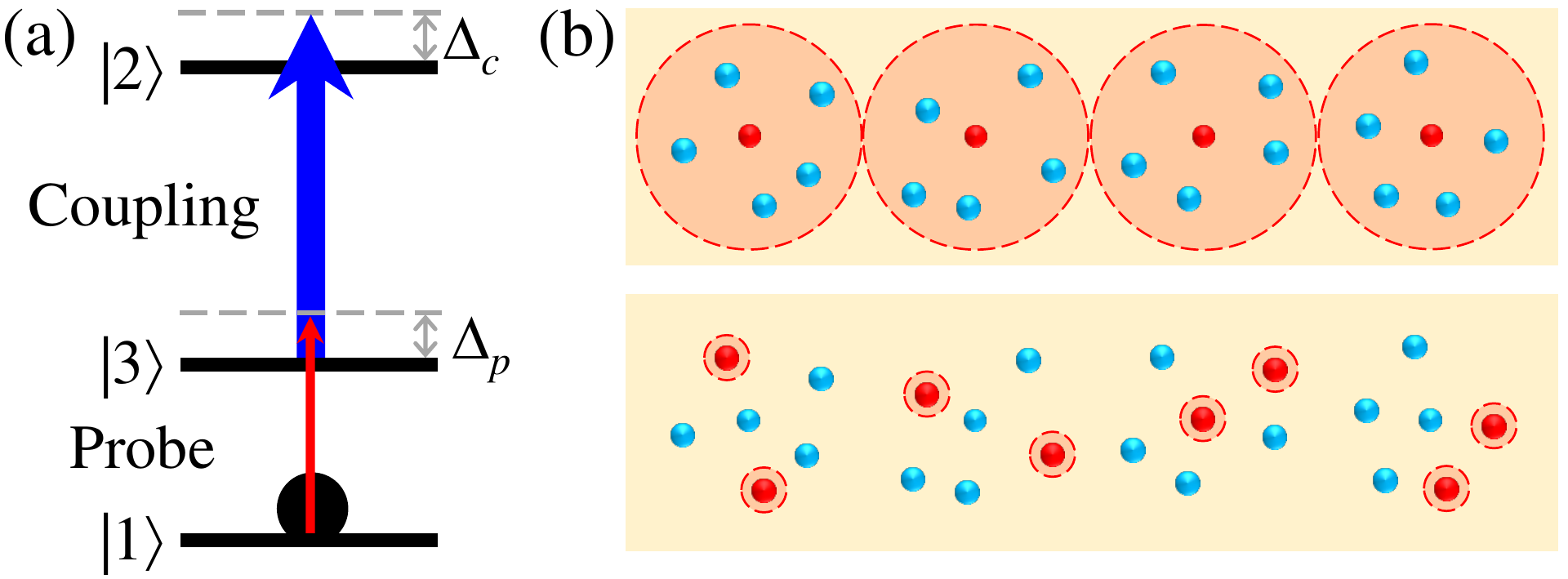}}
	\caption{(a) Transition diagram of the Rydberg-EIT system. $|1\rangle$, $|2\rangle$, and $|3\rangle$ represent the ground, Rydberg, and intermediate states. The weak probe and strong coupling fields form the ladder-type EIT configuration. (b) Top and bottom figures depict the systems of strongly- and weakly-interacting Rydberg polaritons. Red and blue balls represent atoms with and without Rydberg excitations; dashed circles indicate the blockade spheres. The strong- and weak-interaction systems are characterized by $r_B^3/r_a^3 \rightarrow 1$ and $r_B^3/r_a^3 \ll 1$, respectively, where $r_B$ is the blockade radius and $r_a$ is the half mean distance between Rydberg excitations or polaritons. As an example, let us consider 8 photons in both systems. There are 8 Rydberg excitations in the weak-interaction system, but only 4 in the strong-interaction system due to the dipole blockade effect.}
	\label{fig:Transition}
	\end{figure}
}
\newcommand{\FigTwo}{
	\begin{figure}[t]
	\center{\includegraphics[width=0.65\columnwidth]{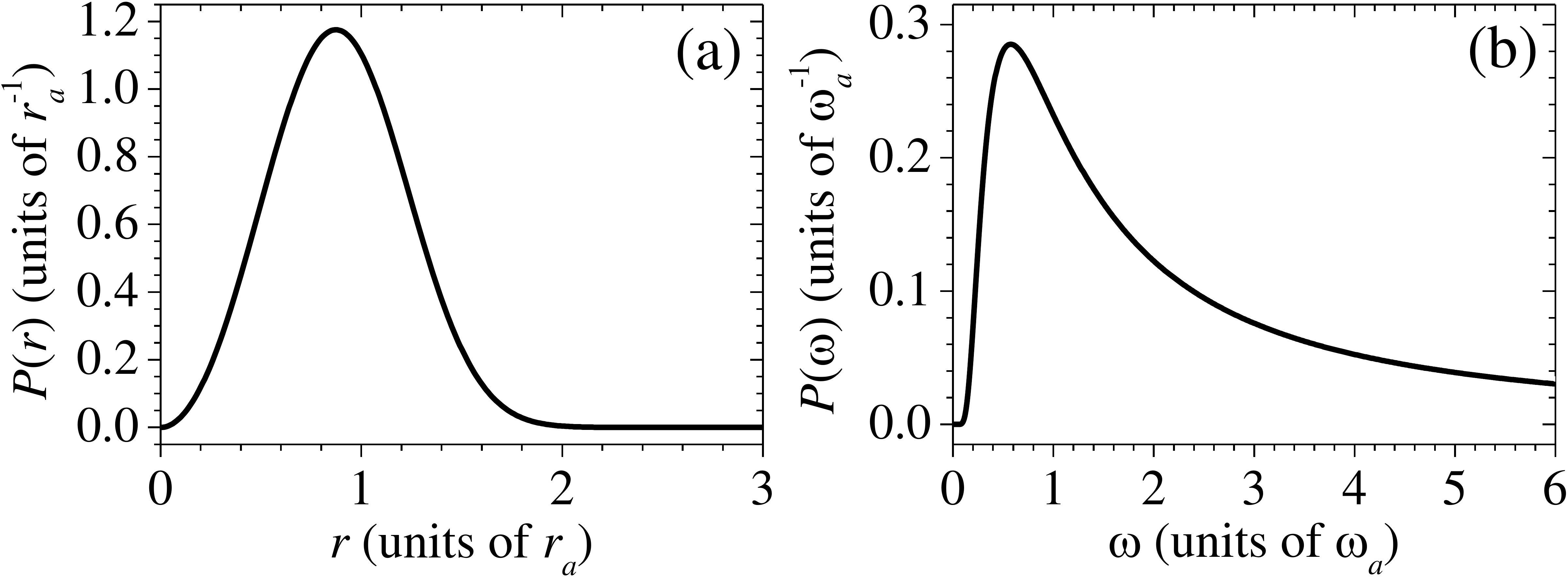}}
	\caption{(a) Probability density $P(r)$ as a function of distance $r$ in the nearest-neighbor distribution. Units of $r_a$ defined by Eq.~(\ref{eq:Define_r_a}) is the half mean distance between particles. (b) Probability density $P(\omega)$ as a function of frequency shift $\omega$. Units of $\omega_a$ defined by Eq.~(\ref{eq:Define_omega_a}) is the frequency shift corresponding to $r_a$.}
	\label{fig:Distribution}
	\end{figure}
}
\newcommand{\FigThree}{
	\begin{figure}[t]
	\center{\includegraphics[width=0.95\columnwidth]{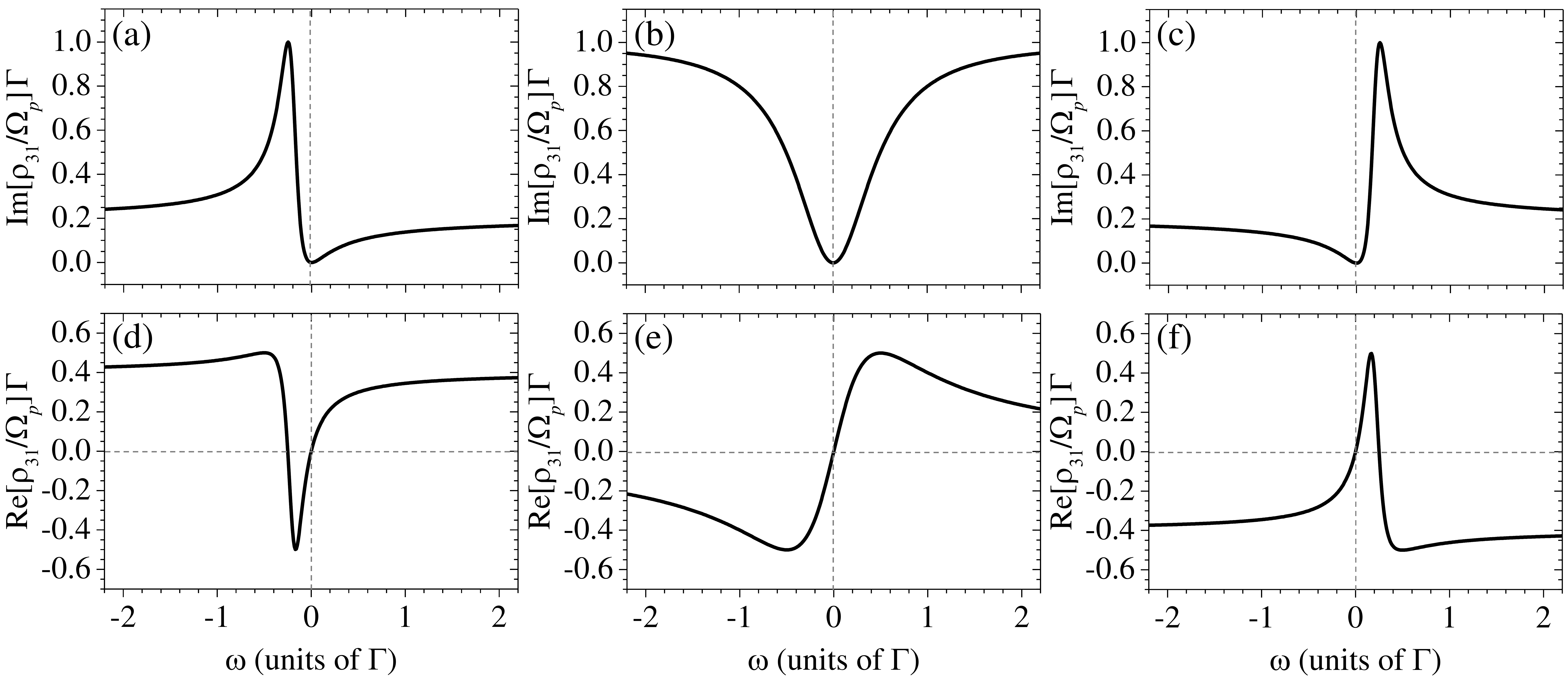}}
	\caption{Imaginary and real parts of $(\rho_{31}/\Omega_p)\Gamma$ as functions of the frequency shift $\omega$ at the two-photon resonance. We calculate the spectra by making the substitutions of first $\Delta_c \rightarrow \Delta_c + \omega$ and then $\Delta_p \rightarrow -\Delta_c$ in Eq.~(\ref{eq:rho31_omegap}) with $\Omega_c = 1.0\Gamma$, $\gamma_0 = 0$, and $\Delta_c = 1.0$$\Gamma$ in (a,d), 0 in (b,e), and $-1.0$$\Gamma$ in (c,f).}
	\label{fig:rho31}
	\end{figure}
}
\newcommand{\FigFour}{
	\begin{figure}[t]
	\center{\includegraphics[width=0.95\columnwidth]{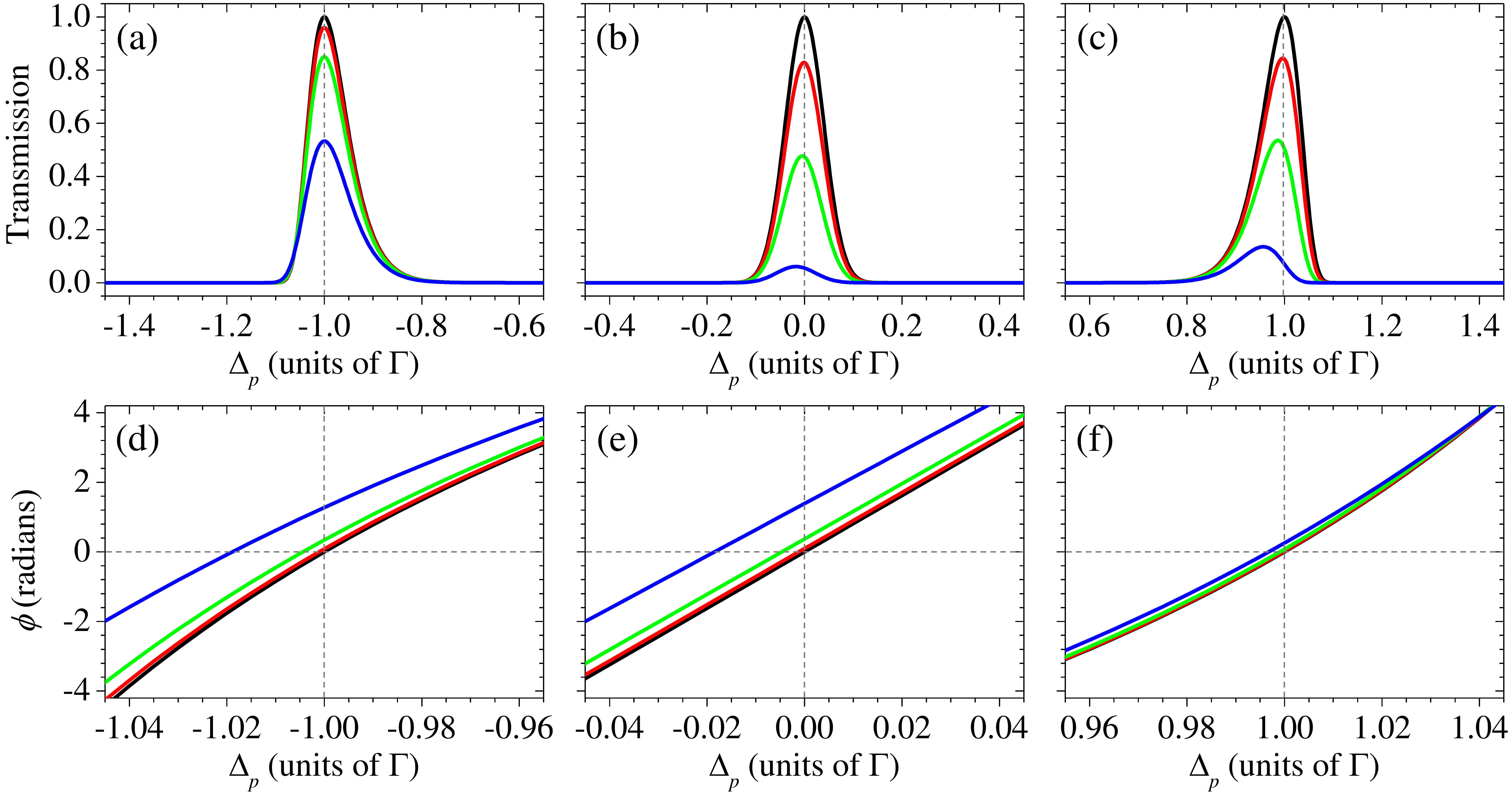}}
	\caption{(a-c) Transmissions of the probe field as functions of the probe detuning, $\Delta_p$. (d-f) Phase shifts of the probe field as functions of $\Delta_p$. The left two, middle two, and right two figures correspond to the coupling detunings, $\Delta_c$, of $+1.0$$\Gamma$, 0, and $-1.0$$\Gamma$, respectively. The vertical axes of the top three figures have the same scale, and those of the bottom three figures also have the same scale. Black lines represent predictions without DDI. Red, green, and blue lines represent predictions with DDI at $ \Omega_{p,\rm{in}} =$ 0.05$\Gamma$, 0.1$\Gamma$, and 0.2$\Gamma$. All the predictions are calculated with $\alpha$ = 81, $\Omega_c$ = 1.0$\Gamma$, and $\gamma_0 = 0$ in Eqs.~(\ref{eq:beta_ddi}) and (\ref{eq:phi_ddi}), and $|C_6|[(4\pi/3)n_{\rm atom}\varepsilon]^2 = 0.35$$\Gamma$ in Eq.~(\ref{eq:omega_a_eit}).
}
	\label{fig:Spectra}
	\end{figure}
}
\newcommand{\FigFive}{
	\begin{figure}[t]
	\center{\includegraphics[width=0.65\columnwidth]{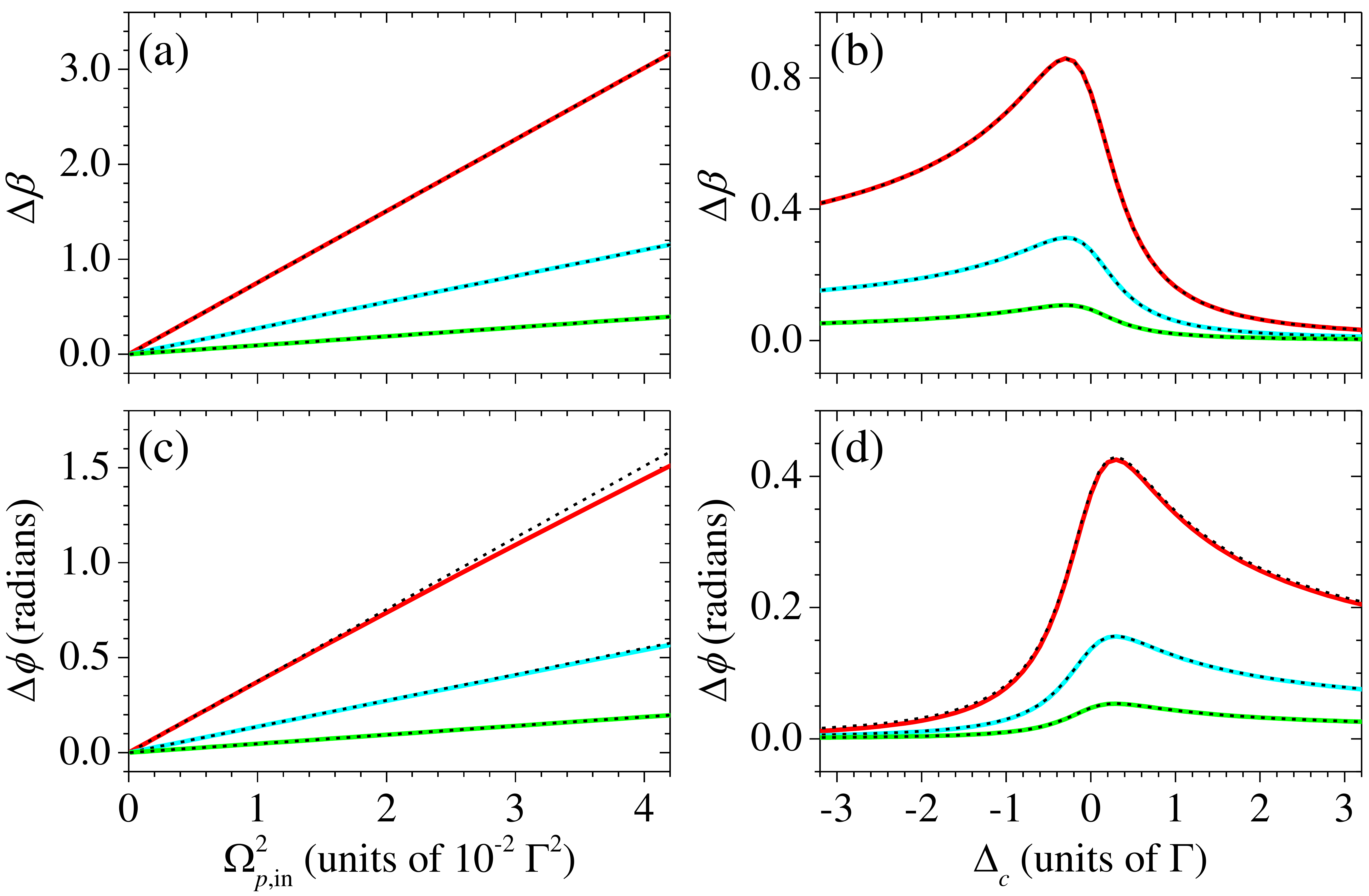}}
	\caption{(a,c) The DDI-induced attenuation coefficient $\Delta \beta$ and phase shift $\Delta \phi$ as functions of $\Omega_{p,\rm{in}}^2$ under $\Delta_p = \Delta_c =0$. (b,d) $\Delta \beta$ and $\Delta \phi$ as functions of $\Delta_c$ under $\delta = 0$ and $\Omega_{p,\rm{in}} = 0.1$$\Gamma$. The horizontal axes of the left two figures have the same scale, so do those of the right two figures. Red, cyan, and green solid lines represent the numerical evaluations of the integrals in Eqs.~(\ref{eq:Delta_beta_DDI_num}) and (\ref{eq:Delta_phi_DDI_num}) at $\Omega_c =$ 1.0$\Gamma$, 1.4$\Gamma$, and 2.0$\Gamma$. Dashed lines are the results of the analytical formulas given by Eqs.~(\ref{eq:Delta_beta_DDI_th_omegap}) and (\ref{eq:Delta_phi_DDI_th_omegap}). All the predictions are calculated with $\alpha$ = 81, $\gamma_0 = 0$, and $|C_6|[(4\pi/3)n_{\rm atom}\varepsilon]^2 = 0.35$$\Gamma$.
}
	\label{fig:DDI_Effect}
	\end{figure}
}
\newcommand{\FigSix}{
	\begin{figure}[t]
	\center{\includegraphics[width=0.65\columnwidth]{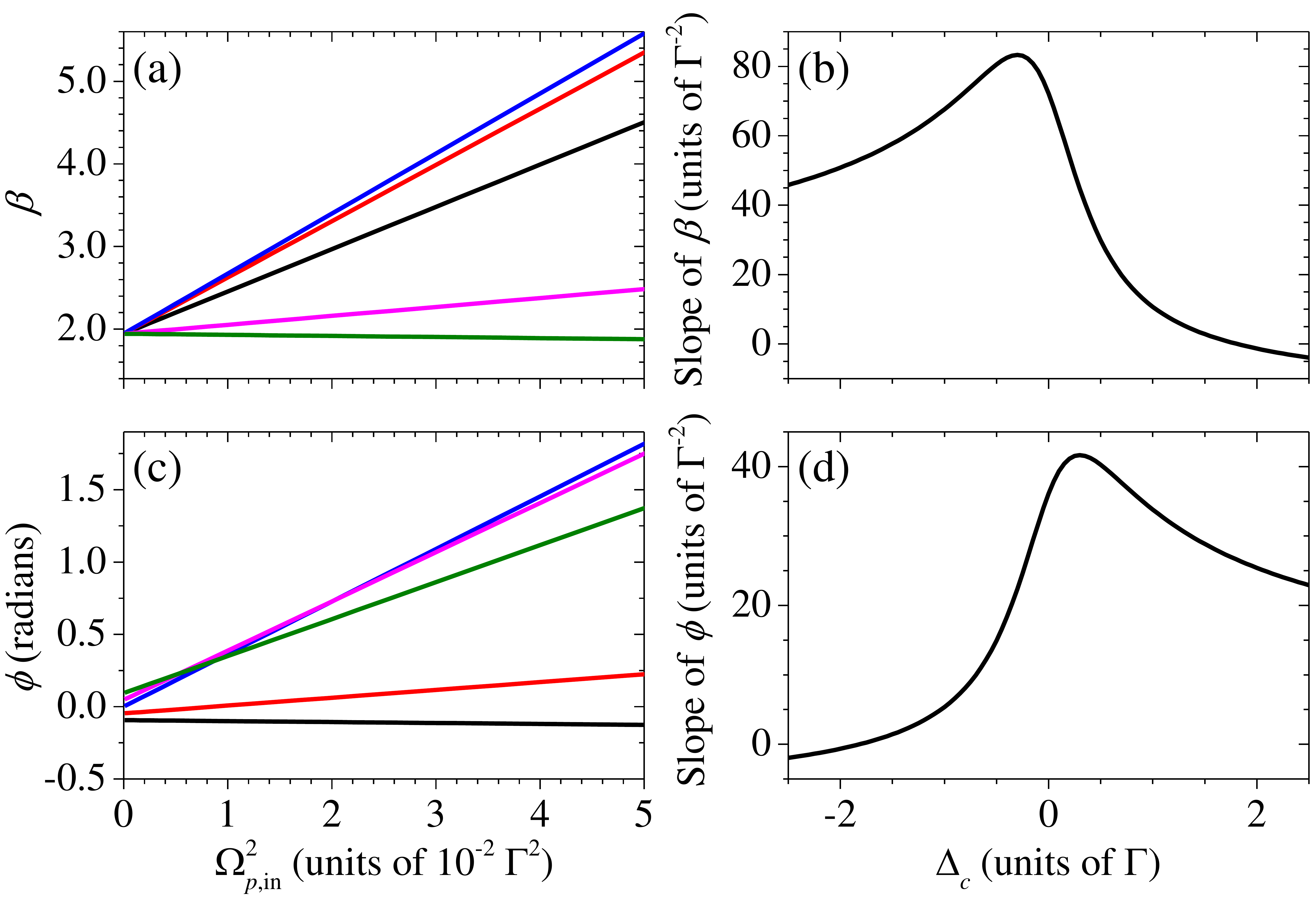}}
	\caption{
Simulation of the experimental data shown in Fig.~2 of Ref.~\cite{OurExp}. In the simulation, $\alpha$ = 81, $\Omega_c =$ 1.0$\Gamma$, $\delta = 0$, $\gamma_0 =$ 0.012$\Gamma$, and $|C_6|[(4\pi/3)n_{\rm atom}\varepsilon]^2 = 0.35$$\Gamma$. (a,c) Attenuation coefficient $\beta$ and phase shift $\phi$ as functions of $\Omega_{p,\rm{in}}^2$ at $\Delta_c = -2$$\Gamma$ (black), $-1$$\Gamma$ (red), 0 (blue), 1$\Gamma$ (magenta), and 2$\Gamma$ (olive). (b,d) Slope of $\beta$ versus $\Omega_{p,\rm{in}}^2$ and that of $\phi$ versus $\Omega_{p,\rm{in}}^2$ as functions of $\Delta_c$.}
	\label{fig:DDI_exp_simulation}
	\end{figure}
}
\newcommand{\FigSeven}{
	\begin{figure}[t]
	\center{\includegraphics[width=0.65\columnwidth]{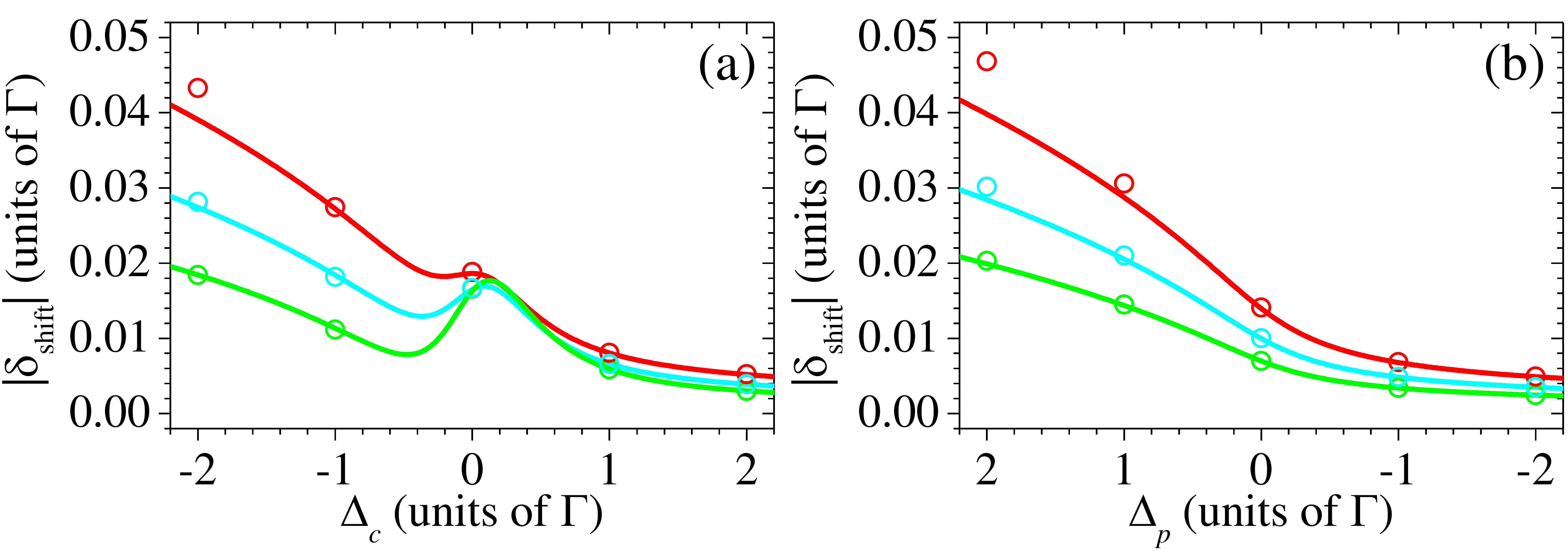}}
	\caption{(a) Magnitude of the DDI-induced EIT peak shift, $\delta_{\rm shift}$, as a function of $\Delta_c$. Solid lines are the predictions of Eq.~(\ref{eq:peak_th}), and circles are those obtained from the spectra numerically calculated by sweeping the probe frequency. (b) Magnitude of $\delta_{\rm shift}$ as a function of $\Delta_p$. Solid lines are the predictions of Eq.~(\ref{eq:peak_th_C}), and circles are those obtained from the spectra numerically calculated  by sweeping the coupling frequency. In (a) and (b), $\Omega_p = 0.2$$\Gamma$, and $\Omega_c = 1.0$$\Gamma$ (red), 1.4$\Gamma$ (cyan), and 2.0$\Gamma$ (green). All the predictions are calculated with $\gamma_0 = 0$ and $|C_6|[(4\pi/3)n_{\rm atom}\varepsilon]^2 = 0.35$$\Gamma$.}
	\label{fig:EIT_peak_shift}
	\end{figure}
}
\section{Introduction}

The effect of electromagnetically induced transparency (EIT) involving Rydberg-state atoms is of great interest currently. The Rydberg-state atoms exhibit the strong electric dipole-dipole interaction (DDI) among each other~\cite{blockade_Zoller2000, blockade_Gould2004, blockade_Pfau2007, SaffmanRMP, blockade_Pohl2013}. On the other hand, the EIT effect not only provides high optical nonlinearity for the atom-light interaction, but also gives rise to slow, stored, and stationary light for long interaction time \cite{EIT_Fleischhauer2005, EIT_OurPRL2006, SLP_OurPRL2012, EIT_YFChen2016, EIT1, EIT2, EIT3, EIT4, EIT5}. Thus, the combination of the strong DDI of Rydberg atoms and the high optical nonlinearity of EIT can efficiently mediate the interaction between photons via Rydberg polaritons in the dipole blockade regime, where the Rydberg polariton is the collective excitation involving the light and the atomic coherence between the ground and Rydberg states~\cite{DSP_Fleischhauer2000, DSP_Fleischhauer2002}. The Rydberg-EIT mechanism can lead to the applications of quantum optics and quantum information processing~\cite{REIT_Adams2010, REIT_Fleischhauer2011, photon_interaction_Lukin2011, REIT_Fleischhauer2015, REIT_Lukin2012, REIT_Hofferberth2016, simulator_Lukin2018, simulator_Lukin2017, SP_transistor_Rempe2014, SP_switch_Rempe2014, SP_transistor_Hofferberth2014, gate_Lukin2019, gate_Rempe2019, SP_Pfau2018, XPM_Rempe2016, Ruseckas2017}. 

To our knowledge, most of the present theoretical models dealt with the Rydberg-EIT system in the strongly-interacting regime, i.e., $r_B^3$ is comparable to $r_a^3$, where $r_B$ is the blockade radius and $r_a$ is the half mean distance between Rydberg polaritons. In Ref.~\cite{REIT_Adams2010}, J.\ D.\ Pritchard {\it et al}.\ utilized the $N$-atom model to analyze experiment phenomena of the optical nonlinearity and attenuation in the Rydberg-EIT system. In Ref.~\cite{REIT_Fleischhauer2011}, D.\ Petrosyan {\it et al}.\ modeled the propagation of light field in strongly-interacting Rydberg-EIT media by considering the superatoms with the volume of the blockade sphere. In Ref.~\cite{photon_interaction_Lukin2011},  A.\ V.\ Gorshkov {\it et al}.\ proposed a theory for the propagation of few-photon pulses in the system of strongly-interacting Rydberg polaritons. In Ref.~\cite{REIT_Fleischhauer2015}, M.\ Moos {\it et al}.\ utilized a one-dimensional model to describe the time evolution of Rydberg polaritons and analyze many-body phenomena in the strongly-interacting regime. In Ref.~\cite{Ruseckas2017},  J.\ Ruseckas {\it et al}.\ proposed a method to create two-photon states by making pairs of Rydberg atoms entangled during the storage.

\FigOne

In this article, we considered the weakly-interacting Rydberg-EIT system, and developed a mean field model to describe the attenuation and phase shift of the output probe field induced by the DDI effect. The Rydberg-EIT system is depicted in Fig.~\ref{fig:Transition}(a), and the weakly-interacting condition requires $r_B^3 \ll r_a^3$ [see Fig.~\ref{fig:Transition}(b)].  Under such condition, the system of Rydberg polaritons can be considered as nearly the ideal gas. Thus, the nearest-neighbor distribution (NND) shown by Ref.~\cite{NNDistribution} is utilized in our model. The DDI-induced frequency shift between nonuniformly-distributed Rydberg excitations results in the effective phase shift and attenuation of light field. With the probability function of NND and the atom-light coupling equations of EIT system, we calculated the mean field results of transmission and phase shift spectra, and further derived the analytical formulas of the DDI-induced attenuation coefficient and phase shift. The theoretical predictions from the formulas are in good agreement with the experimental data in Ref.~\cite{OurExp}. In the experiment of Ref.~\cite{OurExp}, we utilized the Rydberg state of a low principal number, the laser-cooled ensemble of a moderate atomic density, and the weak probe field of a low photon flux to make the mean number of Rydberg polaritons within the blockade sphere lower than 0.1. The good agreement verifies our model. 

Rydberg polaritons are regarded as bosonic quasi-particles, and the DDI-induced phase shift and attenuation coefficient can infer the elastic and inelastic collision rates in the ensemble of these particles \cite{DSP_BEC_Fleischhauer2008}. Weakly-interacting Rydberg polaritons assisted by a long interaction time of the EIT effect can be employed in the study of many-body physics such as the Bose-Einstein condensation of polaritons \cite{EPBEC_nature2006, EPBEC_science2007, DSP_BEC_Fleischhauer2008, EPBEC_RMP2010}. The mean field theory developed in this work provides a useful tool to conceive ideas relevant to weakly-interacting EIT-based Rydberg polaritons and to evaluate feasibilities of experiments.

We organize the article as follows. In Sec.~\ref{sec:2}, the theoretical model based on the probability function of NND, the atom-light coupling equations of the EIT system, and the ensemble average of the DDI-induced frequency shift are introduced. We obtain the mean field results of the real and imaginary parts of the steady-state absorption cross section of the probe field. In Sec.~\ref{sec:3}, we numerically evaluate the integrals corresponding to the mean field results and present the spectra of transmission and phase shift of the output probe field. The DDI-induced phenomena observed from the spectra are discussed and explained. In Sec.~\ref{sec:4}, we derive the analytical formulas of the DDI-induced attenuation coefficient and phase shift. From the formulas, one can see how the DDI effects depend on the system parameters such as the optical depth, coupling and probe Rabi frequencies, coupling detuning, two-photon detuning, and decoherence rate. It is interesting to note that the DDI effects exhibit the asymmetric behavior with respect to the coupling detuning. In Sec.~\ref{sec:5}, we briefly describe the experimental condition and data in Ref.~\cite{OurExp}, and calculate the predictions corresponding to the experimental condition from the analytical formulas. The predictions are in good agreement with the data. Finally, we give a summary in Sec.~\ref{sec:6}.

\section{Theoretical model}
\label{sec:2}

In the system of Rydberg polaritons, the DDI induces a frequency shift of the Rydberg state. Since Rydberg excitations are nonuniformly distributed, the Rydberg-state frequency shift is not a constant in the medium. The system of low-density Rydberg excitations can be considered as nearly the ideal gas, in which the nearest-neighbor distribution (NND) is given by~\cite{NNDistribution}
\begin{equation}
	P(r)=\frac{3r^{2}}{r_{a}^{3}}e^{-r^{3}/r_{a}^{3}},
\label{eq:P(r)}
\end{equation}
where $P(r)$ is the probability density, i.e., $P(r)dr$ is the probability of finding a particle's nearest neighbor locating at the distant between $r$ and $r+dr$, and $r_{a}$ is the half mean distance between particles. The definition of $r_a$ is
\begin{equation}
	r_{a} \equiv \left( \frac{3}{4\pi n_R} \right)^{1/3},
\label{eq:Define_r_a}
\end{equation}
where $n_R$ is the Rydberg-polariton density. Figure~\ref{fig:Distribution}(a) shows $P(r)$ as a function of $r$.

The NND is the consequence of each particle being randomly distributed. In Appendix VII of Ref.~\cite{NNDistribution}, Eqs.~(669)-(671) and the relating descriptions explain how the NND is derived. The derivation is summarized in the following: The probability $P(r)dr$ satisfies the equation of
	$P(r)dr = [(4\pi r^2 dr)n_R]\times \left[ 1-\int_{0}^{r}P(r')dr' \right],$
where $(4 \pi r^2 dr)n_R$ is the probability of finding one particle within the volume of $4 \pi r^2 dr$ under the particle's density $n_R$, and $1-\int_{0}^{r}P(r')dr'$ is that of all the remaining particles locating outside a sphere of the radius $r$. By eliminating $dr$ and taking the derivative on both sides of the equation, we obtain
	$(d/dr) \left[ P(r)/(4 \pi r^2 n_R) \right] = -P(r)$.
The solution of this differential equation gives Eq.~(\ref{eq:P(r)}). Thus, as long as the interaction between the Rydberg excitations is weak enough to maintain the nature of random distribution, Eq. (\ref{eq:P(r)}) is valid for the Rydberg-EIT system.

The frequency shift of a Rydberg state induced by the DDI is $C_6/r^6$, where $C_6$ is the van der Waals coefficient \cite{C6_Saffman2008} and $r$ represents the distance between two particles. In the ensemble of Rydberg excitations, the Rydberg-state frequency shift consists of two parts. The first part is $C_6/r^6$ contributed from the nearest-neighbor Rydberg excitation at the distance $r$, and the second part is $n_R \int_{r}^{\infty} (C_6/r'^6) 4\pi r'^2 dr'$ contributed from all the other Rydberg excitations outside the sphere of the radius $r$. Here, we consider the particles in the second part are uniformly distributed. Thus, the Rydberg-state frequency shift is the following:
\begin{equation}
	\omega=\frac{C_6}{r^6}+\frac{C_6}{r_a^3 r^3}.
\label{eq:omega}
\end{equation}
Using Eqs.~(\ref{eq:P(r)}) and (\ref{eq:omega}), we can obtain frequency shift distribution $P(\omega)$, i.e., $P(\omega)d\omega$ is the probability of finding the Rydberg-state frequency shifted by the  amount between $\omega$ and $\omega+d\omega$, given by
\begin{equation}
\begin{aligned}
	P(\omega) = \frac{1}{\omega_a}
	\frac{\left[ 1+\sqrt{1+4(\omega/\omega_a)} \right]^2}
	{4 (\omega/\omega_a)^2 \sqrt{1+4(\omega/\omega_a)}}
 	\exp{\left[ -\frac{1+\sqrt{1+4(\omega/\omega_a)}}{2(\omega/\omega_a)} \right]}, \\
\end{aligned}	
\label{eq:P(omega)}
\end{equation}
where
\begin{equation}
		\omega_a \equiv |C_6| /r_a^6.
\label{eq:Define_omega_a}
\end{equation}
Since the value of distance, $r$, is always positive, only $\omega \geq 0$ is valid in  $P(\omega)$. Figure~\ref{fig:Distribution}(b) shows $P(\omega)$ as a function of $\omega$.

\FigTwo

In the EIT system shown in Fig.~\ref{fig:Transition}(a), the weak probe field drives the transition between the ground state $|1\rangle$ and the intermediate state $|3\rangle$, and the strong coupling field drives that between $|3\rangle$ and the Rydberg state $|2\rangle$. We consider the steady-state continuous-wave case in this work. As the system reaches its steady state, a given amount of Rydberg excitations with the density $n_R$ are produced and distributed according to Eq.~(\ref{eq:P(r)}). The weak probe field propagates through the system consisting of the atoms with their Rydberg-state levels shifted by the existing Rydberg excitations via the DDI \cite{Phol2011}. We will first use the optical Bloch equation (OBE) to calculate the optical coherence of the probe transition, which determines the susceptibility of the probe field. Then, the susceptibility will be averaged over the frequency shift $\omega$ according to the distribution function in Eq.~(\ref{eq:P(omega)}), producing a factor of $n_R$ in the averaged susceptibility. Because $n_R$ is proportional to the square of the probe-field amplitude or Rabi frequency, this gives rise to the nonlinearity in the system. Finally, we will employ the averaged susceptibility in the Maxwell-Schr\"{o}dinger equation (MSE) to obtain the attenuation and phase shift of the probe field caused by the DDI-shifted Rydberg-state levels.


We utilize the OBE of the atomic density matrix and the MSE of the probe field of the EIT system in the theory. The complete OBE and MSE are shown below, but their time-derivative terms are dropped in the calculation because we consider the steady-state case.
\begin{eqnarray}
\label{eq:OBE_rho21}
	\frac{\partial}{\partial t}\rho_{21} &=& 
		\frac{i}{2}\Omega_{c}\rho_{31} +i\delta\rho_{21}
		-\left( \gamma_0+\frac{\Gamma_2}{2} \right) \rho_{21}, \\
\label{eq:OBE_rho31}		
	\frac{\partial}{\partial t}\rho_{31} &=&
		\frac{i}{2}\Omega_{p} +\frac{i}{2}\Omega_{c}\rho_{21}
		+i\Delta_{p}\rho_{31} -\frac{\Gamma}{2}\rho_{31}, \\
\label{eq:OBE_rho22}
	\frac{\partial}{\partial t}\rho_{22} &=&
		\frac{i}{2}\Omega_{c}\rho_{32}-\frac{i}{2}\Omega_{c}\rho_{32}^*
		-\Gamma_2\rho_{22},\\
\label{eq:OBE_rho32}
	\frac{\partial}{\partial t}\rho_{32} &=&
		\frac{i}{2}\Omega_{p}\rho_{21}^* +\frac{i}{2}\Omega_{c}(\rho_{22}-\rho_{33})
		- \left( \frac{\Gamma_2+\Gamma}{2}+i \Delta_c \right) \rho_{32}, \\
\label{eq:OBE_rho33}
	\frac{\partial}{\partial t}\rho_{33} &=&
		-\frac{i}{2}\Omega_{p}^*\rho_{31} +\frac{i}{2}\Omega_{p}\rho_{31}^*
		-\frac{i}{2}\Omega_{c}\rho_{32} + \frac{i}{2}\Omega_{c}\rho_{32}^* -\Gamma\rho_{33}, \\
\label{eq:MSE}
	\frac{1}{c}\frac{\partial}{\partial t}\Omega_p
		&+& \frac{\partial}{\partial z}\Omega _p = i\frac{\alpha\Gamma}{2L}\rho_{31},
\end{eqnarray}
where $\rho_{ij}$ is the density matrix element between states $|i\rangle$ and $|j\rangle$, $\Omega_{p}$ and $\Omega_{c}$ represent the probe and coupling Rabi frequencies, $\Delta_p$ and $\Delta_c$ are the one-photon detunings of the probe and coupling transitions, $\delta = \Delta_p + \Delta_c$ is the two-photon detuning, $\gamma_0$ is the decoherence or dephasing rate of the Rydberg coherence $\rho_{21}$, $\Gamma$ is the spontaneous decay rate of $|3\rangle$ which is $2\pi$$\times$6 MHz in our case of the state $|5P_{3/2}\rangle$ of $^{87}$Rb atoms, $\Gamma_2$ is the spontaneous decay rate of $|2\rangle$ which is $2\pi$$\times$5.4 kHz in our case of the state $|32D_{5/2}\rangle$, and $\alpha$ and $L$ are the optical depth (OD) and the length of the medium. Since $\Omega_p \ll \Omega_c$ and $\Omega_p \ll \Gamma$ in this work, we treat the probe field as a perturbation and keep only the terms of the lowest order of $\Omega_p$ in each equation. The value of $\Gamma_2$ is small and, thus, we set it to zero throughout this work. 

We will determine the optical coherence, $\rho_{31}$, which is responsible for the attenuation coefficient and phase shift of the probe field. Equations~(\ref{eq:OBE_rho21}) and (\ref{eq:OBE_rho31}) without the time-derivative terms are used to obtain the steady-state solution of $\rho_{31}$ given by
\begin{equation}
	\rho_{31}(\Delta_p , \Delta_c) = 
		\frac{\Delta_p + \Delta_c + i\gamma_0}
		{\Omega_{c}^{2}/2- 2(\Delta_p+ i\Gamma/2)(\Delta_p + \Delta_c + i\gamma_0)}
		\Omega_p.
\label{eq:rho31_omegap}
\end{equation}
With above $\rho_{31}$, we solve Eq.~(\ref{eq:MSE}) and find the ratio of output to input probe Rabi frequencies as the following:
\begin{equation}
	\frac{\Omega_p(L)}{\Omega_p(0)} = \exp(i\phi-\beta/2),
\label{eq:out_Omegap}
\end{equation}
where $\beta$ and $\phi$ represent the attenuation coefficient and phase shift of the probe field at the output, and the probe transmission is $\exp(-\beta)$. We use $\beta_0$ and $\phi_0$ to denote the attenuation coefficient and phase shift without the DDI effect. The optical coherence of the probe field determines $\beta_0$ and $\phi_0$ as the followings:
\begin{eqnarray}
\label{eq:beta0}
	\beta_0 (\Delta_p,\Delta_c) &=& \alpha \Gamma 
		\;{\rm Im}\! \left[ \frac{\rho_{31} (\Delta_p,\Delta_c)}{\Omega_p} \right], \\
\label{eq:phi0}
	\phi_0 (\Delta_p,\Delta_c) &=& \frac{\alpha \Gamma}{2} 
		\,{\rm Re}\! \left[ \frac{\rho_{31} (\Delta_p,\Delta_c)}{\Omega_p} \right].
\end{eqnarray}

The effect of DDI on the attenuation coefficient and phase shift of the probe field will be derived below. Due to the DDI-induced frequency shift of the Rydberg state, the one-photon detuning of the coupling field transition is shifted by the amount of $\omega$, i.e., 
\begin{equation}
	\Delta_c \rightarrow \Delta_c\pm\omega. \nonumber
\end{equation}
Because $\omega \geq 0$, the positive or negative sign in the above corresponds to negative or positive $C_6$, respectively, and we use $+\omega$ which corresponds to negative $C_6$ in the following. Under the DDI, the probe field propagates through the atoms with different DDI-induced frequency shifts, where the probability density $P(\omega)$ of the frequency shift distribution has been shown in Eq.~(\ref{eq:P(omega)}). We obtain the values of $\beta$ and $\phi$ by averaging $\rho_{31}$ over the frequency distribution as shown below.
\begin{equation}
	\beta (\Delta_p,\Delta_c) = \alpha \Gamma \int_{0}^{\infty} d\omega P(\omega)
		\;{\rm Im}\!\left[ \frac{\rho_{31} (\Delta_p,\Delta_c+\omega)}{\Omega_p} \right],
\label{eq:beta_ddi}
\end{equation}
\begin{equation}
	\phi (\Delta_p,\Delta_c)=\frac{\alpha \Gamma}{2} \int_{0}^{\infty} d\omega P(\omega) 
		\;{\rm Re}\! \left[ \frac{\rho_{31} (\Delta_p,\Delta_c+\omega)}{\Omega_p} \right].
\label{eq:phi_ddi}
\end{equation}
To show the effect of the frequency shift, we plot the imaginary and real parts of $(\rho_{31}/\Omega_p)\Gamma$ against $\omega$ at the condition of the two-photon resonance, i.e., $\Delta_p + \Delta_c = 0$, in  Fig.~\ref{fig:rho31}. Please note that the integrals in Eqs.~(\ref{eq:beta_ddi}) and (\ref{eq:phi_ddi}) are carried out from $\omega = 0$ to $\infty$, and thus the integration results of positive and negative one-photon detunings are rather different. 

\FigThree

The dipole blockade effect is that an atom inside the blockade sphere cannot be excited to the Rydberg state, where the blockade sphere centering with a Rydberg excitation has the radius $r_B \equiv (2 C_6 \Gamma/\Omega_c^2)^{1/6}$ \cite{REIT_Lukin2012}. This effect has already been included in the integrals of Eqs.~(\ref{eq:beta_ddi}) and (\ref{eq:phi_ddi}). In the weakly-interacting system considered here, i.e., $r_{B}^3 \ll r_a^3$, the average number of Rydberg excitations per volume of the blockade sphere is far less than one, and thus the dipole blockade appears rarely.

To evaluate Eqs.~(\ref{eq:beta_ddi}) and (\ref{eq:phi_ddi}), one needs to know the value of $\omega_a$ in $P(\omega)$. According to the definition of $\omega_a$ in Eq.~(\ref{eq:Define_omega_a}) and that of $r_a$ in Eq.~(\ref{eq:Define_r_a}), we can relate $\omega_a$ to the Rydberg-polariton density, $n_R$, as $\omega_a = |C_6| [(4\pi/3) n_R]^2$. The product of the atomic density, $n_{\rm atom}$, and the average Rydberg-state population, $\bar{\rho}_{22}$, gives $n_R$, and therefore $\omega_a = |C_6| [(4\pi/3) n_{\rm atom} \bar{\rho}_{22}]^2$. The DDI-induced nonlinear and many-body effects make $\bar{\rho}_{22}$ no longer be the steady-state solution of the OBE shown in Eqs.~(\ref{eq:OBE_rho21})-(\ref{eq:OBE_rho33}). Nevertheless, one can phenomenologically associate $\bar{\rho}_{22}$ to the steady-state solution of Rydberg-state population  at the input, $\rho_{22,{\rm in}}$, by introducing a parameter $\varepsilon$.
Substituting $\varepsilon \rho_{22,{\rm in}}$ for $\bar{\rho}_{22}$, we obtain
\begin{equation}
	\omega_a  = |C_6|  
		\left[ (4\pi/3) n_{\rm atom} \varepsilon \rho_{22,{\rm in}} \right]^2,
\label{eq:omega_a_eit}
\end{equation}
where $\varepsilon$ is the phenomenological parameter representing the average value of entire ensemble.

\section{Predictions of transmission and phase-shift spectra}
\label{sec:3}

The spectra of probe transmission and phase shift under the DDI effect are obtained by numerically evaluating the integrals of Eqs.~(\ref{eq:beta_ddi}) and (\ref{eq:phi_ddi}) with the value of $\omega_a$ given by Eq.~(\ref{eq:omega_a_eit}). Figures~\ref{fig:Spectra}(a)-\ref{fig:Spectra}(c) show the probe transmission versus the probe detuning at the coupling detunings of $+1$$\Gamma$, 0, and $-1$$\Gamma$; similarly, Figs~\ref{fig:Spectra}(d)-\ref{fig:Spectra}(e) show the probe phase shift. The spectra without and with the DDI are calculated with Eq.~(\ref{eq:beta0}) [or Eq.~(\ref{eq:phi0})] and Eq.~(\ref{eq:beta_ddi}) [or Eq.~(\ref{eq:phi_ddi})], respectively. 

The DDI-induced phenomena exhibited in the transmission and phase shift spectra are summarized as follows: (1) A larger probe intensity results in a smaller transmission or larger attenuation. (2) A larger probe intensity results in a larger phase shift. (3) With the same probe intensity, the EIT peak transmission at a positive coupling detuning (e.g., $\Delta_c = +1$$\Gamma$) is larger than that at a negative coupling detuning (e.g., $\Delta_c = -1$$\Gamma$), where the positive and negative detunings have the same magnitude. (4) With the same probe intensity, the phase shift of a positive coupling detuning (e.g., $\Delta_c = +1$$\Gamma$) is larger than that of a negative coupling detuning (e.g., $\Delta_c = -1$$\Gamma$), where the positive and negative detunings have the same magnitude. (5) The position of the EIT peak transmission at $\Delta_c = +1$$\Gamma$ changes very little and locates around $\delta = 0$; that at $\Delta_c = -1$$\Gamma$ shifts away from $\delta = 0$ significantly and a larger probe intensity induces a greater shift. We will explain the five phenomena below.

First of all, the peak transmission decreases against the probe Rabi frequency [see Figs.~\ref{fig:Spectra}(a), \ref{fig:Spectra}(b), and \ref{fig:Spectra}(c)]. This is expected, because the Rydberg-state population is proportional to the probe intensity or Rabi frequency square. A larger Rydberg-state population or Rydberg-polariton density makes $\omega_a$ larger as shown by Eq.~(\ref{eq:omega_a_eit}). The probability density $P(\omega)$ with the larger $\omega_a$ has a broader width and a longer tail as demonstrated by Fig.~\ref{fig:Distribution}(b). Under the broader $P(\omega)$, more atoms have the Rydberg-state frequency shifted away from the EIT resonance condition, reducing the peak transmission more. Secondly, the phase shift increases against the probe Rabi frequency [see Figs.~\ref{fig:Spectra}(d), \ref{fig:Spectra}(e), and \ref{fig:Spectra}(f)]. The explanation is similar to that in the first phenomenon.

\FigFour

The third phenomenon observed in the spectra is the asymmetry in the peak transmissions of Fig.~\ref{fig:Spectra}(a) and Fig.~\ref{fig:Spectra}(c). With the same value of $\Omega_{p,\rm{in}}^2$, the DDI-induced reduction of the peak transmission at $\Delta_c =$ $+1$$\Gamma$ is less than that at $\Delta_c =$ $-1$$\Gamma$. This can be explained with the help of Figs.~\ref{fig:rho31}(a) and 
\ref{fig:rho31}(c), 
which show ${\rm Im}[\rho_{31}/\Omega_p]$ as functions of $\omega$ at the two-photon resonance for $\Delta_c = +1$$\Gamma$ and $-1$$\Gamma$, respectively. To obtain the probe transmission, the integration of Eq.~(\ref{eq:beta_ddi}) is performed only for the region of $\omega >0$. In Fig.~\ref{fig:rho31}(a), the large sharp absorption peak, corresponding to the two-photon transition, in the spectrum of ${\rm Im}[\rho_{31}/\Omega_p]$ locates at the left to $\omega < 0$ and plays no role in the DDI-induced effect. The value of ${\rm Im}[\rho_{31}/\Omega_p]$ is always small for $\omega > 0$, producing a smaller value of $\int_{0}^{\infty} d\omega P(\omega) {\rm Im}[\rho_{31}/\Omega_p]$, i.e., a higher probe transmission. On the other hand, in 
Fig.~\ref{fig:rho31}(c), 
the large sharp absorption peak locates at the right to $\omega$. This peak produces a larger value of $\int_{0}^{\infty} d\omega P(\omega) {\rm Im}[\rho_{31}/\Omega_p]$, and reduces the transmission significantly. Thus, the location of the two-photon-transition peak with respect to $\omega = 0$ in the spectrum of ${\rm Im}[\rho_{31}(\omega)/\Omega_p]$ is responsible for the asymmetry that at the same probe Rabi frequency the peak transmission in Fig.~\ref{fig:Spectra}(a) is larger than that in Fig.~\ref{fig:Spectra}(c).

The fourth phenomenon observed in the spectra is that the probe intensity or $\Omega_{p,\rm{in}}^2$ has a much larger effect on the phase shift of $\delta = 0$ at $\Delta_c =$ 1$\Gamma$ as shown by Fig.~\ref{fig:Spectra}(d) than that at $\Delta_c =$ $-1$$\Gamma$ as shown by Fig.~\ref{fig:Spectra}(f). This can be explained with the help of 
Figs.~\ref{fig:rho31}(d) and \ref{fig:rho31}(f), 
which show ${\rm Re}[\rho_{31}/\Omega_p]$ at $\Delta_c =$ 1$\Gamma$ and $-1$$\Gamma$, respectively. To obtain the phase shift, the integration of Eq.~(\ref{eq:phi_ddi}) is performed only for the region of $\omega >0$. In 
Fig.~\ref{fig:rho31}(d),
the value of ${\rm Re}[\rho_{31}/\Omega_p]$ is always positive for $\omega > 0$, resulting in a larger value of $\int_{0}^{\infty} d\omega P(\omega) {\rm Re}[\rho_{31}/\Omega_p]$, i.e., a larger phase shift. On the other hand, in 
Fig.~\ref{fig:rho31}(f), 
${\rm Re}[\rho_{31}/\Omega_p]$ has both positive and negative values for $\omega > 0$, because the resonance of the two-photon transition locates at $\omega > 0$. The cancellation between positive and negative values of the integrand makes $\int_{0}^{\infty} d\omega P(\omega) {\rm Re}[\rho_{31}/\Omega_p]$ nearly zero, i.e., almost no phase shift. Therefore, with the same value of $\Omega_{p,\rm{in}}$, the phase shift at $\Delta_c =$ 1$\Gamma$ shown by Fig.~\ref{fig:Spectra}(d) is significant, and that at $\Delta_c =$ $-1$$\Gamma$ shown by Fig.~\ref{fig:Spectra}(f) is little. 

The fifth phenomenon is that at $\Delta_c = +1$$\Gamma$ the EIT peak positions of different $\Omega_p$ are all very close to $\delta = 0$ as shown in Fig.~\ref{fig:Spectra}(a), and at $\Delta_c = -1$$\Gamma$ those of $\Omega_p = 0.1$$\Gamma$ and $0.2$$\Gamma$ shift away from $\delta = 0$ significantly as shown in Fig.~\ref{fig:Spectra}(c). Furthermore, in Fig.~\ref{fig:Spectra}(c) a larger value of $\Omega_p$ results in a larger shift. Because of $C_6 < 0$, all the shifts should be negative as expected. Please refer to Figs.~\ref{fig:rho31}(a) and  \ref{fig:rho31}(c) plotted at $\Delta_c = +1$$\Gamma$ and $-1$$\Gamma$, respectively. In each of the two plots, a negative shift (or a smaller value of $\Delta_p$) makes the entire solid line move to the right. In Fig.~\ref{fig:rho31}(a), the magnitude of the shift can only be small, otherwise the big sharp absorption peak moves toward $\omega = 0$, and the result of the integration from $\omega = 0$ to $\infty$ becomes larger, i.e., the probe transmission decreases. On the other hand, in Fig.~\ref{fig:rho31}(c) the magnitude of the shift can be large such that the big sharp absorption peak moves further away from $\omega = 0$, and the result of the integration becomes smaller, i.e., the probe transmission increases. Thus, the magnitude of the shift is asymmetric with respect to the one-photon detuning. In the Appendix, we will derive an analytical formula to quantitatively predict the DDI-induced frequency shift of the EIT peak.

\section{Analytical formulas of the DDI-induced attenuation coefficient and phase shift}
\label{sec:4}

We now derive the analytical formulas for the DDI-induced attenuation coefficient, $\Delta\beta$, and phase shift, $\Delta\phi$, at the condition of $\gamma_0 = 0$ and $\delta = 0$ (or $\Delta_p = -\Delta_c$). Here, $\Delta\beta$ (or $\Delta\phi$) is defined as the difference between the values of $\beta$ (or $\phi$) with and without the DDI effect, i.e., $\Delta\beta\equiv\beta -\beta_0$ and $\Delta\phi\equiv \phi -\phi_0$. 

At $\gamma_0 = 0$ and $\delta = 0$, Eq.~(\ref{eq:rho31_omegap}) gives $\beta_0 = 0$ and $\phi_0 = 0$, and thus $\Delta\beta$ = $\beta$ and $\Delta\phi$ = $\phi$. Replacing $\Delta_p$ by $-\Delta_c$ in $\beta$ of Eq.~(\ref{eq:beta_ddi}) and in $\phi$ of Eq.~(\ref{eq:phi_ddi}), we obtain $\Delta\beta$ and $\Delta\phi$ as follows:
\begin{eqnarray}
\label{eq:Delta_beta_DDI_num}
	\Delta\beta 
	&=& \alpha \Gamma \int_{0}^{\infty} d\omega P(\omega) 
		\frac{4\omega^2\Gamma}{4\omega^2 \Gamma^2+(4\omega\Delta_c+\Omega_c^2)^2}, \\
\label{eq:Delta_phi_DDI_num}
	\Delta\phi 
	&=& \frac{\alpha \Gamma}{2} \int_{0}^{\infty} d\omega P(\omega) 
		\frac{8\omega^2 \Delta_c +2\omega \Omega_c^2}
		{4\omega^2 \Gamma^2+(4\omega\Delta_c+\Omega_c^2)^2}.
\end{eqnarray}
In the weakly-interacting or low-density system, the region of $\omega$ being the order of $\omega_a$ is very near the center of the EIT window, in which ${\rm Im}[\rho_{31}/\Omega_p]$ and ${\rm Re}[\rho_{31}/\Omega_p]$ are nearly zero and contribute to the above two integrals very little. On the other hand, the region of $\omega \gg \omega_a$ is away from the center of the EIT window, and contributes to the above two integrals predominately. Under $\omega \gg \omega_a$, in the integrands of Eqs.~(\ref{eq:Delta_beta_DDI_num}) and (\ref{eq:Delta_phi_DDI_num}) we can make the approximation of $P(\omega)$ as
\begin{equation}
	P(\omega) \approx \frac{\sqrt{\omega_a}}{2 \omega^{3/2}} \equiv P'(\omega),
\label{eq:P(omega)2}
\end{equation}
where $\omega_a$ is given by Eq.~(\ref{eq:omega_a_eit}). In Eq.~(\ref{eq:omega_a_eit}), the steady-state solution of $\rho_{22,{\rm in}}$ is
\begin{equation}
	\rho_{22,{\rm in}} =  \frac{\Omega_{p,{\rm in}}^2 \Omega_c^2}
		{4\delta^2\Gamma^2+(\Omega_c^2-4\delta\Delta_p)^2}
		\approx \frac{\Omega_{p,{\rm in}}^2}{\Omega_c^2}, 
\label{eq:rho22_in}
\end{equation}
where $\delta\Gamma, \delta\Delta_p \ll \Omega_c^2$ is the typical condition in most of the EIT experiments. Without any other approximation, we use $P'(\omega)$ in Eqs.~(\ref{eq:Delta_beta_DDI_num}) and (\ref{eq:Delta_phi_DDI_num}) and replace $\rho_{22,{\rm in}}$ in $\omega_a$ by $\Omega_{p,{\rm in}}^2/\Omega_c^2$ to obtain
\begin{eqnarray}
\label{eq:Delta_beta_DDI_th_omegap}
	\Delta\beta &=& 2 S_{\rm DDI} \sqrt{\frac{W_c-2\Delta_c}{W_c^2}} 
		\Omega_{p,{\rm in}}^2, \\
\label{eq:Delta_phi_DDI_th_omegap}
	\Delta\phi &=& S_{\rm DDI} \sqrt{\frac{W_c+2\Delta_c}{W_c^2}} \Omega_{p,{\rm in}}^2,
\end{eqnarray}
where
\begin{eqnarray}
\label{eq:S_DDI}
	S_{\rm DDI} &\equiv& 
		\frac{\pi^2\alpha\Gamma\sqrt{|C_6|} n_{\rm atom} \varepsilon}{3\Omega_c^3}, \\
	W_c &\equiv& \sqrt{\Gamma^2+4\Delta_c^2}.
\end{eqnarray}
The above results being good approximations imposes the condition that $\omega_a$ is much smaller than the EIT linewidth, $\Delta\omega_{\rm EIT}$, where $\Delta\omega_{\rm EIT} =$ $\Omega_c^2 \sqrt{\Gamma^2+8\Delta_c^2}/(\Gamma^2+4\Delta_c^2)$ derived from the spectrum of ${\rm Im}[\rho_{31}(\omega)]$ at $\delta = 0$. More precisely, the accuracy of the analytical formula of $\Delta\beta$ requires $(\omega_a/\Delta\omega_{\rm EIT})^{3/2} \ll 1$, and that of $\Delta\phi$ requires $(\omega_a/\Delta\omega_{\rm EIT})^{1/2} \ll 1$. 

In Fig.~\ref{fig:DDI_Effect}, we compare the results of the above two analytical formulas with those of the numerical integrations of Eqs.~(\ref{eq:Delta_beta_DDI_num}) and (\ref{eq:Delta_phi_DDI_num}) without the approximation of $P(\omega)$. The agreement between the results of the analytical formulas and numerical integrations is satisfactory except the line of $\Delta\phi$ at $\Omega_c = 1.0$$\Gamma$ in the region of $\Omega_{p,\rm{in}}^2 >$ 0.02$\Gamma^2$. In this region, $(\omega_a/\Delta\omega_{\rm EIT})^{1/2} \ll 1$ is no longer well satisfied, and the deviation between the analytical formula and the numerical integration becomes observable. Figures~\ref{fig:DDI_Effect}(a) and \ref{fig:DDI_Effect}(c) demonstrate that both of $\Delta\beta$ and $\Delta\phi$ are proportional to $\Omega_{p,\rm{in}}^2/\Omega_c^3$. Figure~\ref{fig:DDI_Effect}(b) [or \ref{fig:DDI_Effect}(d)] shows the asymmetric phenomenon that the value of $\Delta\beta$ (or $\Delta\phi$) at the coupling detuning of $|\Delta_c|$ is smaller (or larger) than that at the coupling detuning of $-|\Delta_c|$.

\FigFive

In reality, there exist a nonzero decoherence rate $\gamma_0$ and the two-photon detuning $\delta$ in the system. We need to consider the corrections of $\gamma_0$ and $\delta$ to the analytical formulas. Under the condition of $\Omega_c^2 \gg \gamma_0 \Gamma,\delta\Gamma$, the attenuation coefficient and phase shift without the DDI effect, $\beta_0$ and $\phi_0$, are approximately given by
\begin{eqnarray}
\label{eq:beta_0_correct}
	\beta_0 &\approx &\frac{2\alpha\gamma_0\Gamma}{\Omega_c^2}
		-\frac{16\alpha\gamma_0\delta\Delta_c\Gamma}{\Omega_c^4}, \\
\label{eq:phi_0_correct}
	\phi_0&\approx &\frac{\alpha\Gamma\delta}{\Omega_c^2}
		-\frac{4\alpha\gamma_0\delta\Gamma^2}{\Omega_c^4}
		+\frac{4\alpha(\gamma_0^2-\delta^2)\Delta_c\Gamma}{\Omega_c^4}.
\end{eqnarray}
To derive the DDI-induced attenuation coefficient, $\Delta\beta$, and phase shift, $\Delta\phi$, we first use the replacement of $\Delta_c \rightarrow \Delta_c + \omega$, substitute $\delta$ for $\Delta_p + \Delta_c$, and approximate $\Delta_p$ to $-\Delta_c$ in $\rho_{31}/\Omega_p$ shown by Eq.~(\ref{eq:rho31_omegap}). Then, we expand $\rho_{31}/\Omega_p$ with respect to $\gamma_0$ and $\delta$ under the assumption of $\Omega_c^2/\Gamma \gg \gamma_0, \delta$ to obtain
\begin{eqnarray}
\label{eq:im_rho31_gamma_0}
	{\rm Im}\! \left[\frac{\rho_{31}}{\Omega_p} \right] 
		&=& A_0 +A_1\gamma_0 +A_2\delta +\cdots, \\
\label{eq:re_rho31_gamma_0}
	{\rm Re}\! \left[\frac{\rho_{31}}{\Omega_p} \right]
		&=& B_0 +B_1\gamma_0 +B_2\delta +\cdots,
\end{eqnarray}
where
\begin{subequations}
\begin{eqnarray}
\label{eq:A0}
 	A_0 &=& \frac{4\omega^2\Gamma}
		{4\omega^2\Gamma^2+(4\omega\Delta_c+\Omega_c^2)^2},	\\
\label{eq:A_gamma}
	A_1 &=& 
		\frac{2\Omega_c^2[(4\omega\Delta_c+\Omega_c^2)^2-4\omega^2\Gamma^2]}
		{[4\omega^2\Gamma^2+(4\omega\Delta_c+\Omega_c^2)^2]^2},	\\
\label{eq:A_delta}
	A_2 &=& \frac{8\omega\Gamma\Omega_c^2(4\Delta_c\omega+\Omega_c^2)}
		{[4\omega^2\Gamma^2+(4\omega\Delta_c+\Omega_c^2)^2]^2},
\end{eqnarray}
\end{subequations}
and
\begin{subequations}
\begin{eqnarray}
\label{eq:B0}
	B_0 &=& \frac{8\Delta_c\omega^2+2\omega\Omega_c^2}
		{4\omega^2\Gamma^2+(4\omega\Delta_c+\Omega_c^2)^2},\\	
\label{eq:B_gamma}
	B_1 &=& -\frac{8\omega\Gamma\Omega_c^2(4\Delta_c\omega+\Omega_c^2)}
		{[4\omega^2\Gamma^2+(4\omega\Delta_c+\Omega_c^2)^2]^2},	\\
\label{eq:B_delta}
	B_2 &=&
		\frac{2\Omega_c^2[(4\omega\Delta_c+\Omega_c^2)^2-4\omega^2\Gamma^2]} 
		{[4\omega^2\Gamma^2+(4\omega\Delta_c+\Omega_c^2)^2]^2}.
\end{eqnarray}
\end{subequations}
Next, we evaluate the two integrals of Eqs.~(\ref{eq:beta_ddi}) and (\ref{eq:phi_ddi}) by substituting Eqs.~(\ref{eq:im_rho31_gamma_0}) and (\ref{eq:re_rho31_gamma_0}) for ${\rm Im}[\rho_{31}/\Omega_p]$ and ${\rm Re}[\rho_{31}/\Omega_p]$ in the integrands. Since $\omega_a$ is much less than the EIT linewidth, $P'(\omega)$ shown in Eq.~(\ref{eq:P(omega)2}) can be employed in Eqs.~(\ref{eq:beta_ddi}) and (\ref{eq:phi_ddi}) to replace $P(\omega)$. The results of the two integrals give $\beta$ and $\phi$. Finally, the analytical formulas of $\Delta\beta (= \beta-\beta_0)$ and $\Delta\phi (= \phi-\phi_0)$, including the corrections of $\gamma_0$ and $\delta$ are given by
\begin{eqnarray}
\label{eq:beta_DDI_th_correct}
	\Delta\beta &=& 2 S_{\rm DDI}\Bigg( \sqrt{\frac{W_c-2\Delta_c}{W_c^2}}
		-\frac{3\gamma_0\sqrt{W_c +2\Delta_c}}{\Omega_c^2} 
		+\frac{3\delta\sqrt{W_c-2\Delta_c}}{\Omega_c^2} \Bigg)  \Omega_{p,{\rm in}}^2, \\
\label{eq:phi_DDI_th_correct}
	\Delta\phi &=& S_{\rm DDI}\Bigg( \sqrt{\frac{W_c+2\Delta_c}{W_c^2}}
		-\frac{3\gamma_0\sqrt{W_c-2\Delta_c}}{\Omega_c^2} 
		-\frac{3\delta\sqrt{W_c+2\Delta_c}}{\Omega_c^2} \Bigg) \Omega_{p,{\rm in}}^2,	
\end{eqnarray}
Because $\Delta_p$ is approximated as $-\Delta_c$ in the derivation, it is more precise that $\Delta_c$ in Eqs.~(\ref{eq:beta_DDI_th_correct}) and (\ref{eq:phi_DDI_th_correct}) is replaced by $-\Delta_p$ (i.e., $\Delta_c - \delta$). The above two formulas are for $C_6 < 0$. We can make the substitutions of $\Delta\phi \rightarrow -\Delta\phi$, $\Delta_c \rightarrow -\Delta_c$, and $\delta \rightarrow -\delta$ to obtain the formulas for $C_6 > 0$. Regarding $\Delta\beta$ as a function of $\Omega_{p,\rm{in}}^2$ in Fig.~\ref{fig:DDI_Effect}(a), the slope will decrease a little due to $\gamma_0$, and become a little larger (or smaller) due to positive (or negative) $\delta$. Regarding $\Delta\phi$ as a function of $\Omega_{p,\rm{in}}^2$ in Fig.~\ref{fig:DDI_Effect}(c), the slope will decrease a little due to $\gamma_0$, and become a little smaller (or larger) due to positive (or negative) $\delta$. When we consider $\beta$ and $\phi$ instead of $\Delta\beta$ and $\Delta\phi$ in Figs.~\ref{fig:DDI_Effect}(a) and \ref{fig:DDI_Effect}(c), $\beta_0$ and $\phi_0$ make nonzero vertical-axis interceptions of those lines.

\section{Simulation of the experimental data}
\label{sec:5}

To verify the mean field theory developed in this work, we systematically measured the attenuation coefficient, $\beta$, and phase shift, $\phi$, of the output probe field as shown in Fig.~2 of Ref.~\cite{OurExp}. The experiment was carried out in cold $^{87}$Rb atoms with the temperature of 350 $\mu$K. The ground state $|1\rangle$, Rydberg state $|2\rangle$, and excited state $|3\rangle$ in the EIT system here correspond to $|5S_{1/2}, F=2, m_F=2\rangle$, $|32D_{5/2}, m_J=5/2\rangle$, and $|5P_{3/2}, F=3, m_F=3\rangle$ in the experiment. 
We set $\Omega_c = 1.0\Gamma$ and the Rydberg state has $C_6 = -2\pi$$\times$260~MHz$\cdot\mu$m$^6$. The values of $\Omega_c$ and $C_6$ result in $r_B^3 \approx 9.3 \mu m^3$. Furthermore, the atomic density $n_{\rm atom}$ was about 0.05 $\mu$m$^{-3}$ and $\Omega_{p,{\rm in}} \leq 0.2\Gamma$. The values of $n_{\rm atom}$, $\Omega_c$, and $\Omega_{p,{\rm in}}$ give $r_a^3 \geq 120$ $\mu$m$^3$. Thus, $r_B^3/r_a^3 \leq 0.08$, showing that the Rydberg polaritons are weakly-interacting in the experiment. With a given Rydberg state, a low value of $n_{\rm atom} \Omega_{p,{\rm in}}^2/\Omega_c^3$ make Rydberg polaritons weakly interacting. Nevertheless, to observe the DDI effect in the weak-interaction regime, a high OD is the necessary condition. The OD of the cold atom cloud was about 81 in the experiment. 
Other experimental details can be found in Ref.~\cite{OurExp}.

\FigSix

In Fig.~\ref{fig:DDI_exp_simulation}, we made the predictions with Eqs.~(\ref{eq:beta_0_correct}), (\ref{eq:phi_0_correct}), (\ref{eq:beta_DDI_th_correct}), and (\ref{eq:phi_DDI_th_correct}) for the comparison with the experimental data in Fig.~2 of Ref.~\cite{OurExp}. The calculation parameters of OD, coupling Rabi frequency, two-photon detuning, and decoherence rate were determined experimentally. 
As for the value of $\sqrt{|C_6|} n_{\rm atom} \varepsilon$ used in the calculation, $n_{\rm atom}$ mentioned above is estimated from the experimental condition, and $\varepsilon$ is determined by fitting the experimental data of the slopes of $\beta$ and $\phi$ versus $\Delta_c$. In the fitting, $\varepsilon$ is the only fitting parameter and the best fits give $\varepsilon = 0.43$. 
Figures~\ref{fig:DDI_exp_simulation}(a) and \ref{fig:DDI_exp_simulation}(c) show the attenuation coefficient, $\beta$, and phase shift, $\phi$, of the output probe field as functions of $\Omega_{p,\rm{in}}^2$,
where $\Omega_{p,\rm{in}}$ (and also $\Omega_c$) is the Rabi frequency at the center of the input Gaussian beam in the experiment. In the derivation of Eqs.~(\ref{eq:beta_DDI_th_correct}) and (\ref{eq:phi_DDI_th_correct}), we do not consider the Gaussian intensity profiles of the probe and coupling fields. Nevertheless, as shown in Eqs.~(\ref{eq:omega_a_eit}) and (\ref{eq:rho22_in}) the phenomenological parameter $\varepsilon$ relates the average Rydberg-state population $\bar{\rho}_{22}$ in the medium to the value of $\Omega_{p,{\rm in}}^2/\Omega_c^2$. The parameter $\varepsilon$ can account for the correction factor for the effect of nonuniform intensity profiles of the light fields and that of decay of the probe field in the medium.
Figure~\ref{fig:DDI_exp_simulation}(b) [or \ref{fig:DDI_exp_simulation}(d)] shows the slope of the straight line of $\beta$ (or $\phi$) versus $\Omega_{p,{\rm in}}^2$ as a function of $\Delta_c$. Note that the decoherence rate, $\gamma_0$, of 0.012$\Gamma$ makes the $y$-axis interception, i.e., $\beta_0$ or $\phi_0$, becomes nonzero according to Eqs.~(\ref{eq:beta_0_correct}) and (\ref{eq:phi_0_correct}), and changes the slopes very little according to Eqs.~(\ref{eq:beta_DDI_th_correct}) and (\ref{eq:phi_DDI_th_correct}). 

In Fig.~2 of Ref.~\cite{OurExp}, the circles are the experimental data and the lines are their best fits. One can clearly observe the important characteristics of asymmetry in the data of slope versus $\Delta_c$. The consistency between the theoretical predictions in Fig.~\ref{fig:DDI_exp_simulation} here and the experimental data in Fig.~2 of Ref.~\cite{OurExp} is satisfactory. The discrepancies in the $y$-axis interceptions of straight lines between the predictions and best fits are minor, and can be explained by the uncertainties or fluctuations of $\delta$ and $\gamma_0$ in the experiment. Therefore, the mean field theory developed in this work is confirmed by the experimental data.

\section{Conclusion}
\label{sec:6}

In summary, a mean field theory based on the nearest-neighbor distribution is developed to describe the DDI effect in the system of weakly-interacting EIT-Rydberg polaritons. 
We deal with the steady-state continuous-wave case in this work. As the system driven by the probe and coupling fields reaches its steady state, Rydberg excitations or polaritons of a given density are produced and locate randomly as described by the nearest-neighbor distribution in Eq.~(\ref{eq:P(r)}). The probe field propagates through the system consisting of the atoms with their Rydberg-state levels shifted by the already existing Rydberg excitations via the DDI. We calculate the optical coherence $\rho_{31}$ of the probe transition, and average $\rho_{31}$ over the frequency shift $\omega$ according to the probability density function of $\omega$ in Eq.~(\ref{eq:P(omega)}). The averaged $\rho_{31}$ determines the attenuation and phase shift of the probe field caused by the shifted Rydberg-state levels.
The numerically-calculated spectra of probe transmission and phase shift are shown in Fig.~\ref{fig:Spectra}. We explain the DDI-induced phenomena observed from the spectra. To make the theory convenient for predicting experimental outcomes and evaluating experimental feasibility, analytical formulas of the DDI-induced attenuation coefficient, $\Delta \beta$, and phase shift, $\Delta \phi$, are derived. As long as $\omega_a$ is much smaller than the EIT linewidth, the results of analytical formulas are in good agreement with those of numerical calculations. According to the formulas, $\Delta \beta$ and $\Delta \phi$ are linearly proportional to $\Omega_{p,\rm{in}}^2$ as demonstrated in Fig.~\ref{fig:DDI_Effect}(a) and \ref{fig:DDI_Effect}(c), and $\Delta \beta$ and $\Delta \phi$ as functions of $\Delta_c$ are asymmetric with respect to $\Delta_c = 0$ as demonstrated in Fig.~\ref{fig:DDI_Effect}(b) and \ref{fig:DDI_Effect}(d). We further consider the existences of nonzero but small decoherence rate and two-photon detuning in the system, and make corrections to the formulas of $\Delta \beta$ and $\Delta \phi$ as shown in Eqs.~(\ref{eq:beta_DDI_th_correct}) and (\ref{eq:phi_DDI_th_correct}). Finally, we make the predictions with the parameters determined experimentally and compare them with the experimental data in Ref.~\cite{OurExp}. The good agreement between the predictions and data demonstrates the validity of our theory. 
Here the steady-state density of Rydberg polaritons is given in the present method, and we have not investigated the transient evolution of Rydberg-polariton density. The theoretical method for the study of nonlinear dynamics of Rydberg polaritons, such as transient behavior and pulse propagation, can be referred to Ref.~\cite{Optica2019}.
Rydberg polaritons are regarded as bosonic quasi-particles, and the DDI is the origin of the interaction between the particles. Thus, the DDI-induced phase shift and attenuation coefficient can infer the elastic and inelastic collision rates in the ensemble of these bosonic particles. Our mean field theory provides a useful tool for conceiving ideas relevant to the EIT system of weakly-interacting Rydberg polaritons, and for evaluating experimental feasibility.

\section*{Appendix}

The DDI effect shifts the position of the EIT peak in the transmission spectrum a little at $\Delta_c = +1$$\Gamma$ as shown by Fig.~\ref{fig:Spectra}(a) and significantly at $\Delta_c = -1$$\Gamma$ as shown by Fig.~\ref{fig:Spectra}(c). We will derive an analytical formula to quantitatively predict the DDI-induced frequency shift of the EIT peak in this Appendix.

We start with $\rho_{31}/\Omega_p$ in Eq.~(\ref{eq:rho31_omegap}). Since we are interested in the EIT peak position but not transmission, $\gamma_0 = 0$ is used in the derivation for simplicity and without sacrificing the generality. The DDI-induced frequency shift of a Rydberg state results in the replacement of $\Delta_c \rightarrow \Delta_c + \omega$ in Eq.~(\ref{eq:rho31_omegap}).
The spectra in Fig.~\ref{fig:Spectra} are obtained by sweeping the probe frequency at a given coupling detuning. Thus, $\Delta_p$ is expressed by $-\Delta_c + \delta$ in Eq.~(\ref{eq:rho31_omegap}), and $\Delta_c$ is treated as a fixed parameter. Under the condition of $\delta \ll \Delta \omega_{\rm EIT}$, we expand  ${\rm Im}[\rho_{31}/\Omega_p]$ with respect to $\delta$ as the followings:
\begin{equation}
	{\rm Im}\left[\frac{\rho_{31}}{\Omega_p}\right] \approx 
		C_0 + C_1 \delta + C_2 \delta^2,
\label{eq:rho31_series}
\end{equation}
where
\begin{subequations}
\begin{eqnarray}
\label{eq:C_0}
	C_0 &=& \frac{4\omega^2\Gamma}
		{4\omega^2\Gamma^2+(\Omega_c^2+4\Delta_c\omega)^2}, \\ 	
\label{eq:C_1}
	C_1 &=& \frac{
		8\Gamma(16\omega^4\Delta_c+4\omega^3\Omega_c^2+4\omega^2\Delta_c\Omega_c^2
		+\omega\Omega_c^4)}
		{[4\omega^2\Gamma^2+(\Omega_c^2+4\Delta_c\omega)^2]^2}, \\
\label{eq:C_2}
	C_2 &=& \frac{4 \Gamma}{\Omega_c^4}[1 + \xi(\omega, \Delta_c, \Omega_c)],
\end{eqnarray}
\end{subequations}
where $\xi$ is a complicate function of $\omega$, $\Delta_c$, and $\Omega_c$. Because $(\omega_a/\Delta\omega_{\rm EIT})^{1/2} \ll 1$ is satisfied in the weak-interaction regime, the contribution of $\xi$ to the integration $\int d\omega P(\omega) C_2$ is negligible, and we can drop $\xi$ from the derivation of the analytical formula of $\delta_{\rm shift}$ for simplicity. Next, we average ${\rm Im}[\rho_{31}/\Omega_p]$ over $\omega$ with the NND and obtain 
$\int d\omega P(\omega) {\rm Im}\left[ \rho_{31}/\Omega_p \right]
	\approx \left[ \int d\omega P(\omega) C_0 \right] 
	+\left[ \int d\omega P(\omega) C_1 \right] \delta 
	+\left[ \int d\omega P(\omega) C_2 \right] \delta^2$.
With the DDI effect, the EIT peak position shifts to $\delta_{\rm shift}$, which minimizes $\int d\omega P(\omega) {\rm Im}\left[ \rho_{31}/\Omega_p \right]$. Since $\int d\omega P(\omega) {\rm Im}\left[ \rho_{31}/\Omega_p \right]$ is a quadratic function of $\delta$, its minimum locates at
\begin{equation}
	\delta_{\rm shift} =-\frac{\int d\omega P(\omega) C_1}{2 \int d\omega P(\omega)C_2}.
\label{eq:peak_num}
\end{equation}
Because $C_2$ is independent of $\omega$ after $\xi$ is dropped, the evaluation of $\int d\omega P(\omega) C_2$ gives $C_2$. In the evaluation of $\int d\omega P(\omega) C_1$, we approximate $P(\omega)$ as $P'(\omega)$ of Eq.~(\ref{eq:P(omega)2}) to obtain an analytical expression. Finally, the frequency shift of the EIT peak is given by
\begin{equation}
	\delta_{\rm shift} = 
		-\frac{\pi^2\sqrt{|C_6|} n_{\rm atom} \varepsilon}{12\Gamma\Omega_c}
		\left[ 3\sqrt{W_c-2\Delta_c}
		+ \left( 2\Delta_c\sqrt{W_c-2\Delta_c} +\Gamma\sqrt{W_c+2\Delta_c} \right)
		\frac{\Omega_c^2}{W_c^3} \right] 
		\Omega_{p,{\rm in}}^2.
\label{eq:peak_th}
\end{equation}
As expected, the frequency shift of the EIT peak is always negative due to $C_6 < 0$. We can make the substitutions of $\delta_{\rm shift} \rightarrow -\delta_{\rm shift}$ and $\Delta_c \rightarrow -\Delta_c$ to obtain the formula for $C_6 > 0$.

It can be seen from Eq.~(\ref{eq:peak_th}) that the magnitude of $\delta_{\rm shift}$ is linearly proportional to $\Omega_{p,{\rm in}}^2$ and independent of the optical depth $\alpha$. Since $W_c \equiv \sqrt{\Gamma^2 + 4\Delta_c^2}$, the result of $\delta_{\rm shift}$ as a function of $\Delta_c$ shows the magnitude of $\delta_{\rm shift}$ at $+|\Delta_c|$ is less than that at $-|\Delta_c|$ as long as $|\Delta_c| \geq \Omega_c/2$. This is consistent with the phenomena shown in Figs.~\ref{fig:Spectra}(a) and \ref{fig:Spectra}(c) that the frequency shift of the EIT peak at $\Delta_c = +1.0$$\Gamma$ is a little and that at  $\Delta_c = -1.0$$\Gamma$ is significant. Figure~\ref{fig:EIT_peak_shift}(a) compares $|\delta_{\rm shift}|$ predicted by Eq.~(\ref{eq:peak_th}) with that determined  from the numerically-calculated spectrum. As long as the condition of $|\delta_{\rm shift}| \ll \Delta \omega_{\rm EIT}$ is satisfied, the analytical formula is in the good agreement with the numerical result.

\FigSeven

The Rydberg-EIT spectra can also be obtained by sweeping the coupling frequency at a given probe detuning. An analytical formula for the DDI-induced EIT peak shift in such spectra is useful. We derive the formula by using $\rho_{31}/\Omega_p$ of Eq.~(\ref{eq:rho31_omegap}) again. In Eq.~(\ref{eq:rho31_omegap}), $\Delta_c$ is expressed by $-\Delta_p + \delta$ and $\Delta_p$ is treated as a fixed parameter. We expand ${\rm Im}[\rho_{31}/\Omega_p]$ with respect to $\delta$, and follow the similar procedure in the paragraph consisting of Eq.~(\ref{eq:peak_th}). Finally, the frequency shift of the EIT peak is given by
\begin{eqnarray}
\label{eq:peak_th_C}
	\delta_{\rm shift} &=& 
		-\frac{\pi^2\sqrt{|C_6|} n_{\rm atom} \varepsilon}{4\Gamma\Omega_c}
		\sqrt{W_p+2\Delta_p} \; 
		\Omega_{p,{\rm in}}^2, \\
	W_p &\equiv& \sqrt{\Gamma^2+4\Delta_p^2}.
\end{eqnarray}
The behavior of Eq.~(\ref{eq:peak_th_C}) is similar to that of Eq.~(\ref{eq:peak_th}), except that the dependence of $\Delta_p$ in Eq.~(\ref{eq:peak_th_C}) quantitatively differs from that of $\Delta_c$ in Eq.~(\ref{eq:peak_th}). Figure~\ref{fig:EIT_peak_shift}(b) compares $|\delta_{\rm shift}|$ predicted by Eq.~(\ref{eq:peak_th_C}) with that determined from the numerically-calculated spectra. Degrees of consistency between the analytical predictions and the numerical results in Fig.~\ref{fig:EIT_peak_shift}(b) are similar to those in Fig.~\ref{fig:EIT_peak_shift}(a).

\section*{Acknowledgments}
This work was supported by Grant Nos.~107-2745-M-007-001 and 108-2639-M-007-001-ASP of the Ministry of Science and Technology, Taiwan.

\section*{Disclosures}
The authors declare no conflicts of interest.



\begin{thebibliography}{99}
\bibitem{blockade_Zoller2000}
	M. D. Lukin, M. Fleischhauer, R. Cote, L. M. Duan, D. Jaksch, J. I. Cirac, and 
	P. Zoller, ``Dipole Blockade and Quantum Information Processing in Mesoscopic Atomic 
	Ensembles,'' Phys. Rev. Lett. {\bf 87}, 037901 (2001).
\bibitem{blockade_Gould2004}
	D. Tong, S. M. Farooqi, J. Stanojevic, S. Krishnan, Y. P. Zhang, R. C\^{o}t\'{e}, E. E. Eyler, and P. L. Gould, ``Local Blockade of Rydberg Excitation in an Ultracold Gas,'' 
	Phys. Rev. Lett. {\bf 93}, 063001 (2004).		
\bibitem{blockade_Pfau2007}
	R. Heidemann, U. Raitzsch, V. Bendkowsky, B. Butscher, R. L\"{o}w, L. Santos, and 
	T. Pfau, ``Evidence for Coherent Collective Rydberg Excitation in the Strong 
	Blockade Regime,'' Phys. Rev. Lett. {\bf 99}, 163601 (2007).
\bibitem{SaffmanRMP}
	M. Saffman, T. G. Walker, and K. M{\o}lmer, ``Quantum information with Rydberg 
	atoms,'' Rev. Mod. Phys. {\bf 82}, 2313-2363 (2010).
\bibitem{blockade_Pohl2013}
	G. Bannasch, T. C. Killian, and T. Pohl, ``Strongly Coupled Plasmas via 
	Rydberg Blockade of Cold Atoms,'' Phys. Rev. Lett. {\bf 110}, 253003 (2013).
\bibitem{EIT_Fleischhauer2005}	
	M. Fleischhauer, A. Imamoglu, and J. P. Marangos, 
	``Electromagnetically induced transparency: Optics in coherent media,'' 
	Rev. Mod. Phys. {\bf 77}, 633-673 (2005).
\bibitem{EIT_OurPRL2006}
	Y.-F. Chen, C.-Y. Wang, S.-H. Wang, and I. A. Yu, 
	``Low-Light-Level Cross-Phase-Modulation Based on Stored Light Pulses,''  
	Phys. Rev. Lett. {\bf 96}, 043603 (2006).
\bibitem{EIT1}
	Z. B. Wang, K.-P. Marzlin, and B. C. Sanders, 
	``Large Cross-Phase Modulation between Slow Copropagating Weak Pulses in $^{87}$Rb,''  
	Phys. Rev. Lett. {\bf 97}, 063901 (2006).
\bibitem{EIT2}
	S. J. Li, X. D. Yang, X. M. Cao, C. H. Zhang, C. D. Xie, and H. Wang, 
	``Enhanced Cross-Phase Modulation Based on a Double Electromagnetically Induced 
	Transparency in a Four-Level Tripod Atomic System,'' 
	Phys. Rev. Lett. {\bf 101}, 073602 (2008).
\bibitem{EIT3}
	B.-W. Shiau, M.-C. Wu, C.-C. Lin, and Y.-C. Chen, 
	``Low-Light-Level Cross-Phase Modulation with Double Slow Light Pulses,'' 
	Phys. Rev. Lett. {\bf 106}, 193006 (2011).
\bibitem{EIT4}
	V. Venkataraman, K. Saha, and A. L. Gaeta, 
	``Phase modulation at the few-photon level for weak-nonlinearity-based quantum 
	computing,'' Nat. Photonics {\bf 7}, 138-141 (2013).
\bibitem{SLP_OurPRL2012}
	Y.-H. Chen, M.-J. Lee, W. Hung, Y.-C. Chen, Y.-F. Chen, and I. A. Yu, 
	``Demonstration of the Interaction between Two Stopped Light Pulses,'' 
	Phys. Rev. Lett. {\bf 108}, 173603 (2012).
\bibitem{EIT5}
	A. Feizpour, M. Hallaji, G. Dmochowski, and A. M. Steinberg, 
	``Observation of the nonlinear phase shift due to single post-selected photons,'' 
	Nat. Phys. {\bf 11}, 905-909 (2015).
\bibitem{EIT_YFChen2016}
	Z.-Y. Liu, Y- H. Chen, Y.-C. Chen, H.-Y. Lo, P.-J. Tsai, I. A. Yu, Y.-C. Chen, 
	and Y.-F. Chen, ``Large Cross-Phase Modulations at the Few-Photon Level,''  
	Phys. Rev. Lett. {\bf 117}, 203601 (2016).
\bibitem{DSP_Fleischhauer2000}	
	M. Fleischhauer and M. D. Lukin, ``Dark-State Polaritons in Electromagnetically 
	Induced Transparency,'' Phys. Rev. Lett. {\bf 84}, 5094-5097 (2000).
\bibitem{DSP_Fleischhauer2002}
	M. Fleischhauer and M. D. Lukin, ``Quantum memory for photons: Dark-state 
	 polaritons,'' Phys. Rev. A {\bf 65}, 022314 (2002).	
\bibitem{REIT_Adams2010}	
	J. D. Pritchard, D. Maxwell, A. Gauguet, K. J. Weatherill, M. P. A. Jones, 
	and C. S. Adams, ``Cooperative Atom-Light Interaction in a Blockaded Rydberg 
	 Ensemble,'' Phys. Rev. Lett. {\bf 105}, 193603 (2010).
\bibitem{REIT_Fleischhauer2011}	
	D. Petrosyan, J. Otterbach, and M. Fleischhauer,
	``Electromagnetically Induced Transparency with Rydberg Atoms,'' 
	Phys. Rev. Lett. {\bf 107}, 213601 (2011).
\bibitem{photon_interaction_Lukin2011}
	A. V. Gorshkov, J. Otterbach, M. Fleischhauer, T. Pohl, and M. D. Lukin, 
	``Photon-Photon Interactions via Rydberg Blockade,'' 
	Phys. Rev. Lett. {\bf 107}, 133602 (2011).
\bibitem{REIT_Lukin2012}	
	T. Peyronel, O. Firstenberg, Q.-Y. Liang, S. Hofferberth, A. V. Gorshkov, T. Pohl, 
	M. D. Lukin, and V. Vuleti\'{c}, ``Quantum nonlinear optics with single photons 
	enabled by strongly interacting atoms,'' Nature (London) {\bf 488}, 57-60 (2012).
\bibitem{SP_switch_Rempe2014}
	S. Baur, D. Tiarks, G. Rempe, and S. D\"urr, 
	``Single-Photon Switch Based on Rydberg Blockade,'' 
	Phys. Rev. Lett. {\bf112}, 073901 (2014).
\bibitem{SP_transistor_Hofferberth2014}
	H. Gorniaczyk, C. Tresp, J. Schmidt, H. Fedder, and S. Hofferberth, 
	``Single-Photon Transistor Mediated by Interstate Rydberg Interactions,'' 
	Phys. Rev. Lett. {\bf 113}, 053601 (2014).
\bibitem{SP_transistor_Rempe2014}
	D. Tiarks, S. Baur, K. Schneider, S. D\"urr, and G. Rempe, 
	``Single-Photon Transistor Using a F\"orster Resonance,'' 
	Phys. Rev. Lett. {\bf 113}, 053602 (2014).
\bibitem{REIT_Fleischhauer2015}
	M. Moos, M. H\"{o}ning, R. Unanyan, and M. Fleischhauer, 
	``Many-body physics of Rydberg dark-state polaritons in the strongly 
	interacting regime,'' Phys. Rev. A {\bf 92}, 053846 (2015).	
\bibitem{XPM_Rempe2016}	
	D. Tiarks, S. Schmidt, G. Rempe, and S. D\"urr, 
	``Optical $\pi$ phase shift created with a single-photon pulse,'' 
	Sci. Adv. {\bf 2}, e1600036 (2016).
\bibitem{REIT_Hofferberth2016}	
	O. Firstenberg, C. S. Adams, and S. Hofferberth, 
	``Nonlinear quantum optics mediated by Rydberg interactions,'' 
	J. Phys. B {\bf 49}, 152003 (2016).
\bibitem{simulator_Lukin2017}
	H. Bernien, S. Schwartz, A. Keesling, H. Levine, A. Omran, H. Pichler, S. Choi, A. S. 
	Zibrov, M. Endres, M. Greiner, V. Vuleti\'{c}, and M. D. Lukin, 
	``Probing many-body dynamics on a 51-atom quantum simulator,'' 
	Nature (London) {\bf 551}, 579-584 (2017).
\bibitem{Ruseckas2017}
	J. Ruseckas, I. A. Yu, and G. Juzeli\={u}nas, 
	``Creation of two-photon states via interaction between Rydberg atoms during 
	light storage,'' 
	Phys. Rev. A {\bf 95}, 023807 (2017).
\bibitem{simulator_Lukin2018}	
	H. Levine, A. Keesling, A. Omran, H. Bernien, S. Schwartz, A. S. Zibrov, M. Endres, M. Greiner, V. Vuleti\'{c}, and M. D. Lukin, 
	``High-Fidelity Control and Entanglement of Rydberg-Atom Qubits,'' 
	Phys. Rev. Lett. {\bf 121}, 123603 (2018).
\bibitem{SP_Pfau2018}
	F. Ripka, H. K\"{u}bler, R. L\"{o}w, and T. Pfau, 
	``A room-temperature single-photon source based on strongly interacting Rydberg 
	atoms,'' Science {\bf 362}, 446-449 (2018).
\bibitem{gate_Lukin2019}
	H. Levine, A. Keesling, G. Semeghini, A. Omran, T. T. Wang, S. Ebadi, H. Bernien, M. Greiner, V. Vuleti\'{c}, H. Pichler, and M. D. Lukin, 
	``Parallel Implementation of High-Fidelity Multiqubit Gates with Neutral Atoms,'' 
	Phys. Rev. Lett. {\bf 123}, 170503 (2019).	
\bibitem{gate_Rempe2019}
	D. Tiarks, S. Schmidt-Eberle, T. Stolz, G. Rempe, and S. D\"urr, 
	``A photon-photon quantum gate based on Rydberg interactions,'' 
	Nat. Phys. {\bf 15}, 124-126 (2019).
\bibitem{NNDistribution}
	S. Chandrasekhar,
	``Stochastic problems in physics and astronomy,'' 
	Rev. Mod. Phys. {\bf 15}, 1-89 (1943).	
\bibitem{OurExp}	
	B. Kim, K.-T. Chen, S.-S. Hsiao, S.-Y. Wang, K.-B. Li, J. Ruseckas, G. Juzeli\={u}nas, T. Kirova, M. Auzinsh, Y.-C. Chen, Y.-F. Chen, and I. A. Yu,
	``A Weakly-Interacting Many-Body System of Rydberg Polaritons Based on Electromagnetically Induced Transparency,''
	arXiv:2006.13526. 
\bibitem{DSP_BEC_Fleischhauer2008}	
	M. Fleischhauer, J. Otterbach, and R. G. Unanyan, 
	``Bose-Einstein Condensation of Stationary-Light Polaritons,'' 
	Phys. Rev. Lett. {\bf 101}, 163601 (2008).
\bibitem{EPBEC_nature2006}
	J. Kasprzak, M. Richard, S. Kundermann, A. Baas, P. Jeambrun, J. M. J. Keeling, F. M. Marchetti, M. H. Szyma\'{n}ska, R. Andr\'{e}, J. L. Staehli, V. Savona, P. B. Littlewood, B. Deveaud, and L. S. Dang, 
	``Bose-Einstein condensation of exciton polaritons,'' 
	Nature (London) {\bf 443}, 409-414 (2006).
\bibitem{EPBEC_science2007}	
	R. Balili, V. Hartwell, D. Snoke, L. Pfeiffer, and K. West, 
	``Bose-Einstein Condensation of Microcavity Polaritons in a Trap,'' 
	Science {\bf 316}, 1007-1010 (2007).
\bibitem{EPBEC_RMP2010}
	H. Deng, H. Haug, and Y. Yamamoto, 
	``Exciton-Polariton Bose-Einstein Condensation,'' 
	Rev. Mod. Phys. {\bf 82}, 1489-1537 (2010).
\bibitem{C6_Saffman2008}
	T. G. Walker and M. Saffman, 
	``Consequences of Zeeman degeneracy for the van der Waals blockade between Rydberg atoms,'' 
	Phys. Rev. A {\bf 77}, 032723 (2008).
\bibitem{Phol2011}
 	S. Sevin\c{c}li, N. Henkel, C. Ates, and T. Pohl,
	``Nonlocal Nonlinear Optics in Cold Rydberg Gases,''
	Phys. Rev. Lett. {\bf 107}, 153001 (2011).
\bibitem{Optica2019}
	Z. Bai, W. Li, and G. Huang,
	``Stable single light bullets and vortices and their active control in cold Rydberg gases,''
	Optica {\bf 6}, 309-317 (2019).
\end{thebibliography}
\end{document}